# Design, Dynamics, and Dissipation of a Torsional-Magnetic Spring Mechanism

Ali Kanj[1], Rhinithaa P. Thanalakshme[2], Chengzhang Li[3], John Kulikowski[4], Gaurav Bahl[5], Sameh Tawfick[6]

*Abstract*— *We present an analytical and experimental study of torsional magnetic mechanism where the restoring torque is due to magnetic field interactions between rotating and fixed permanent magnets. The oscillator consists of a ball bearing-supported permanent magnet, called the rotor, placed between two fixed permanent magnets called the stators. Perturbing the rotor from its equilibrium angle induces a restoring magnetic torque whose effect is modeled as a torsional spring. This restoring effect is accompanied by dissipation mechanisms arising from structural viscoelasticity, air and electromagnetic damping, as well as friction in the ball bearings. To investigate the system's dynamics, we constructed an experimental setup capable of mechanical, electrical and magnetic measurements. For various rotor-stator gaps in this setup, we validated an analytical model that assumes viscous and dry (Coulomb) damping during the rotor's free response. Moreover, we forced the rotor by a neighboring electromagnetic coil into high amplitude oscillations. We observed unusual resonator's nonlinearity: at large rotor-stator gaps, the oscillations are softening; at reduced gaps, the oscillations stiffen-then-soften. The developed reduced-order models capture the nonlinear effects of the rotor-to-stator and the rotor-to-coil distances. These magnetic oscillators are promising in low-frequency electromagnetic signal transmission and in designing magneto-elastic metamaterials with tailorable nonlinearity.*

## 1. Introduction

There is continued interest in electromechanical systems consisting of oscillating permanent magnets (PMs) due to their high energy density, which can be converted to mechanical power. For instance, new metamaterials take advantage of embedded magnetic particles or oscillating permanent magnetic components to manipulate wave propagation. Such metamaterial structures exhibit distinctive phononic behaviors to achieve frequency conversion of acoustic waves [1], dispersion tunability [2] [3], tailored nonlinearity [4] [5], nonreciprocal propagation [6] [7], and mechanical topological insulation [8] [9]. Origami metamaterials use PMs to enable stabilization [10] and actuation [11] by the magnetic force between the flat panels. In addition to these emerging applications, magnetic materials can be used for vibration damping [12]. Eddy currents – generated by a PM in a nearby moving electric conductor – serve as an active dissipation mechanism that outperforms mechanical brakes [13] [14].

Moreover, rare-earth magnets typically possess large surface magnetic flux densities (0.3-0.6 T), thus offering a promising low-cost method to produce detectable magnetic field at large distances in comparison with magnetic current loops (whose field intensity scales with the delivered electrical current) [15] [16] [17]. Efficient magnetic field production enables very- and ultra-low frequency (VLF and ULF) electromagnetic (EM) transmitters that would leverage under-water and under-ground communication [15]. In fact, different V/ULF transmitter designs are proposed in literature and are based on PM motion for EM field transmission [15] [16] [17] [18] [19] [20] [21] [22]. Hence, it is important to understand and quantitatively model the dynamics of magneto-mechanical oscillators, and to capture the various sources of dissipation that practically arise in these systems.

---

[1]A. Kanj is with the department of Mechanical Science and Engineering, University of Illinois at Urbana-Champaign, Urbana, IL 61801 USA (e-mail: alimk2@illinois.edu).
[2]R. P. Thanalakshme was with the University of Illinois at Urbana-Champaign, Urbana, IL 61801 USA (e-mail: rp10@illinois.edu).
[3]C. Li is with the department of Mechanical Science and Engineering, University of Illinois at Urbana-Champaign, Urbana, IL 61801 USA (e-mail: cl61@illinois.edu).
[4]J. Kulikowski was with the University of Illinois at Urbana-Champaign, Urbana, IL 61801 USA (e-mail: kulkwsk2@stanford.edu).
[5]G. Bahl is with the department of Mechanical Science and Engineering, University of Illinois at Urbana-Champaign, Urbana, IL 61801 USA (e-mail: bahl@illinois.edu).
[6]S. Tawfick is with the department of Mechanical Science and Engineering, University of Illinois at Urbana-Champaign, Urbana, IL 61801 USA (e-mail: tawfick@illinois.edu).

PM-based dynamic mechanical systems exhibit intriguing phenomena whose study requires understanding at the mechanical, magnetic and electrical levels [1] - [22]. In fact, intriguing nonlinear responses are typically observed due to the magnetic field distribution in many of the systems stated earlier due to the field geometry [12] [23]. In essence, the magnetic field configuration can be tailored to achieve tailored nonlinear stiffness and enable some of the desired performances (e.g. phononic wave nonreciprocity). Further, various forms of dissipation play critical roles in the dynamics of magnetic oscillators. PM oscillations near electrical conductors or/and other PMs generate electromagnetic viscous damping (sometimes called core losses, eddy currents damping or magnetic hysteresis) [13] [14] [21]. These viscous dissipations add up to existing structural dissipations resulting from the viscoelastic materials in the structure. Ball bearings used as rotational support [20] [21] induce dry friction forces that introduce further nonlinearities to the system's dynamics [6]. In addition to considering these forces affecting the natural (free) response, the actuation force exerted by an electric coil [6] [22] is modeled by accounting for its role in converting the received electrical input into mechanical output to the PM via the coupling magnetic field medium.

The paper starts by describing, in section 2, the magnetic resonator system design and instrumentation, followed by the derivation of the equations of motion. A device consisting of a free-to-rotate PM (rotor) located between two fixed PMs (stators) is mathematically modeled in section 3. In section 4, the free response dynamics of the rotor in ringdown experiments are studied with an emphasis on the importance of including damping nonlinearity resulting from dry friction that is applied by the rotor's bearings. Section 5 considers the rotor's electromechanical actuation by modeling and identifying the nonlinear electromechanical coupling mechanisms between the rotor and the driving coil. In section 6, the experimental forced frequency response of the rotor using the coil is presented and compared to simulations of the coupled nonlinear electromechanical model developed based on the individually identified models of previous sections. Finally, section 7 concludes the work with a summary of the results and directions for future work.

## 2. System Design

*Oscillator Working Principle*

In this work, we study the system in Fig. 1 that consists of a rotational magnetic oscillator driven by an electromagnetic coil. In this system, the free-to-rotate permanent magnet (PM), called the rotor, oscillates due to its location between two fixed permanent magnets, called the stators. These stators are magnetized along the same direction (cf. Fig. 1a) with which the rotor aligns in the absence of other magnetic fields. This angular position of the rotor designates the equilibrium position about which the perturbed rotor oscillates with the angular displacements $\theta$ of Fig. 1a.

For instance, rotating the oscillator from its equilibrium position induces a restoring torque $T_{Mag}$ exerted by the stators (Fig. 1a), which is modeled as a torsional magnetic spring. In order to drive the systems' oscillations, we use a coil that is placed next to the rotor in a way to generate its central magnetic field orthogonally to the rotor magnetic dipole moment at the equilibrium position (Fig. 1a). This coil magnetic field induces a drive torque $T_{Coil}$ that perturbs the rotor from its equilibrium position.

*Design of the Magnetic Oscillator*

Fig. 1b-d show the realized device of the magnetic oscillator in this study. This design uses N45 grade Neodymium (NdFeB) rectangular-block magnets bought from K&J Magnetics, Inc.

We assembled the rotor, displayed in Fig. 1b, using a 50.8 mm x 12.7 mm x 12.7 mm (2 in x 0.5 in x 0.5 in) block magnet whose ends are fitted to machined titanium endcaps. The endcaps hold the magnet to titanium shafts from each side. The two end shafts are equipped with R188-2RS full Zirconia ($ZrO_2$) ceramic ball bearings from VXB ball bearings. The bearings' inner diameter is 6.35 mm (1/4 in), outer diameter is 12.7 mm (1/2 in) and thickness is 4.76 mm (3/16 in). The rotor assembly suspends vertically by mounting its bearings to machined holes in the top and bottom covers of the device frame as shown in Fig. 1c and 1d.

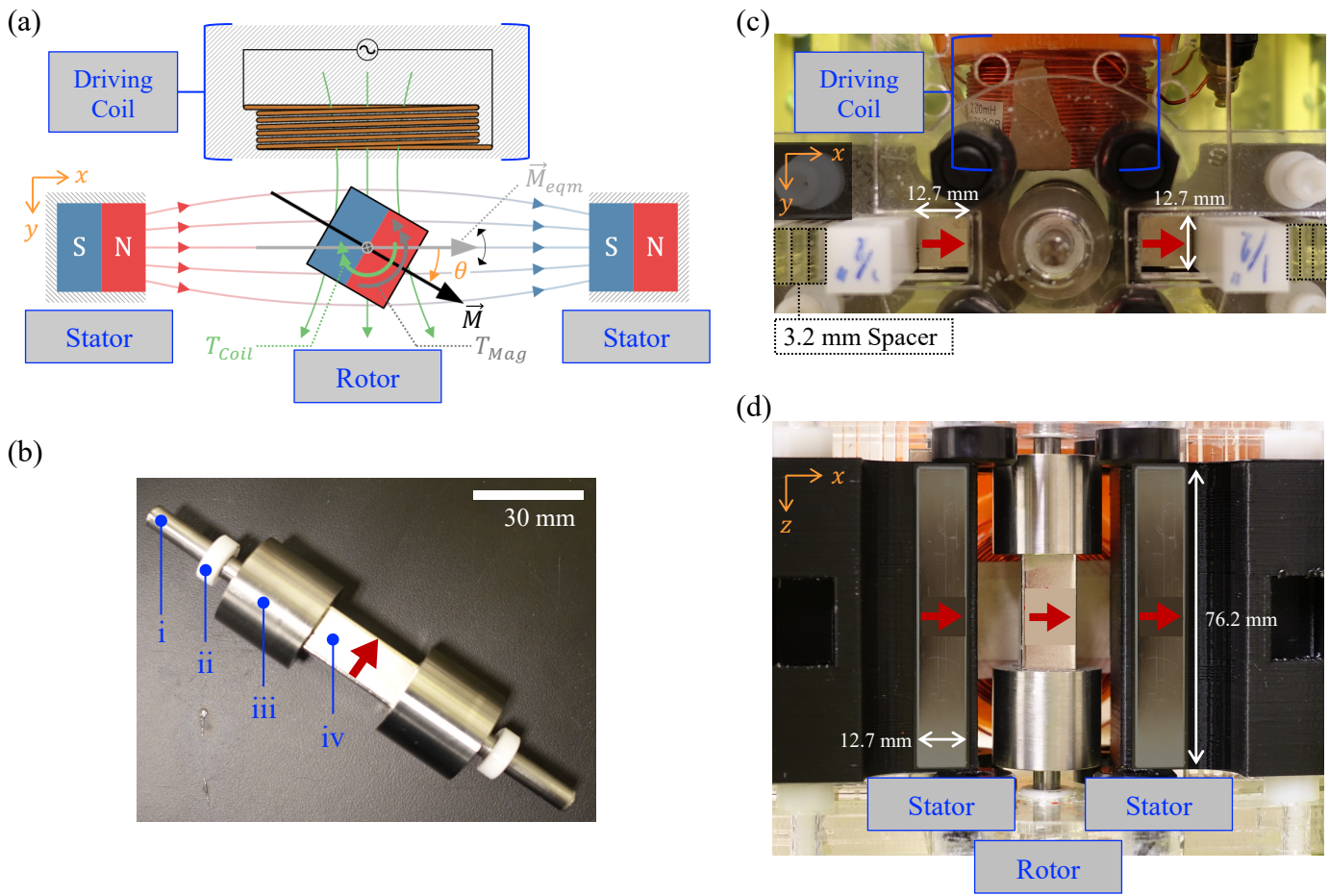

**Fig. 1.** Proposed torsional magnetic spring oscillator. (a) Schematic depicting the concept of the magnetic torsional spring and the coil electromechanical actuation. (b) The rotor assembly with i-titanium shaft, ii-ceramic ball bearings, iii-titanium endcaps, and iv-N45 permanent magnet. (c) Top and (d) side views of the oscillator device used for the experimental study.

These two covers of the frame, which are laser-cut from 12.7 mm thick acrylic sheets, sandwich the 3D-printed PLA+ black stators' housings of Fig 1c and 1d. In this housing, two identical 76.2 mm x 12.7 mm x 12.7 mm (3 in x 0.5 in x 0.5 in) block Neodymium N45 grade magnets serve as stators. In this device, we control the stator-to-rotor distance by varying the number of 3.175 mm (1/8 in) acrylic-spacers inserted between the stator and its housing edge (Fig. 1c), which enables to experimentally study the effect of stator-to-rotor distance on the system's dynamics in sections 3, 4, and 6.

Next to this magnetic arrangement, a 3D-printed PLA+ support holds the 2 mH 18 AWG air-core inductor (Fig. 1c) used as driving coil. This coil support system provides an experimental control of the coil-to-rotor distance with a 6.35 mm (1/4 in) steps to identify the effect of the coil-to-rotor distance on the system's dynamics as explained in section 5.

Complementary to this device, we use a set of instruments to actuate and measure the torque, angles, magnetic field, electric voltages, and electric currents in order to study the different dynamics. We provide in the supplemental material section S1 a detailed description of these instruments and the corresponding measurements.

## 3. Torsional Magnetic Spring

We start by experimentally measuring the magnetic spring stiffness and the effect of stator-to-rotor distance on the torsional oscillator (cf. S1 and supporting video 1). For every stator-to-rotor distance considered, the torque ($T_{Mag}$) exerted by the stators on the rotor is measured for rotor angles $\theta \in [-135°, 135°]$ as plotted in Fig. 2a. To model the dependence of $T_{Mag}$ on $\theta$, we assume that the rotor PM is equivalent to magnetic dipole moment ($\vec{m}_R$) leading to [24]:

$$\vec{T}_{Mag} = \vec{m}_R \times \vec{B}_S = -(m_R B_S \sin\theta)\hat{k} \Longrightarrow T_{Mag} = -\underbrace{m_R B_S}_{T_{amp}} \sin\theta. \tag{1}$$

Equation (1) makes use of the Cartesian coordinate system $(x, y, z)$ of Fig. 1 with associated unit vectors $\hat{\imath}$, $\hat{\jmath}$ and $\hat{k}$ respectively. Equation (1) assumes $\vec{m}_R$ to be located at the rotor center where the stators induce a magnetic field $\vec{B}_S = B_S \hat{\imath}$ oriented along the rotor equilibrium position as shown in Fig. 1a. Treating the rotor as a magnetic dipole moment $\vec{m}_R$ is a valid assumption only when the rotor-stator separation is large [24]. Although for the studied device the dipole approximation might be invalid, this approximation is used to provide an intuitive explanation of the rotor behavior. Hence, referring to (1), the measured values of $T_{Mag}$ over the large angular displacement range are fitted to the nonlinear sinusoidal model of (2) as shown in Fig. 2a. As can be seen by Fig. 2a, the sinusoidal model of amplitude $T_{amp}$ agrees well with the experimental data except for the smallest $d_{Stator} = 2.8$ cm which exhibits small deviations.

To further investigate the magnetic stiffness for small $\theta$, we approximate $\sin\theta$ in (1) with $\theta$ leading to:

$$T_{Mag} = -K_{Mag} \times \theta. \tag{2}$$

We only considered the linearized model in (2) for measurements of small $\theta \in [-45°, 45°]$ as shown in the inset of Fig. 2a. In the inset plots, we see the linearized fit in (2) agrees with the magnetic spring response for these small $\theta$.

To visualize the effect of $d_{Stator}$ on the magnetic stiffness, we present in Fig. 2b the values of linear ($K_{Mag}$) and nonlinear ($T_{amp}$) torsional stiffnesses resulting from fitting of the quasi-static measurements at the respective $d_{Stator}$. The expected equality in values between $K_{Mag}$ and $T_{amp}$ is confirmed by Fig. 2b where the maximum relative error at $d_{Stator} \approx 2.8$ cm does not exceed 16% of the corresponding $K_{Mag}$.

To explain the scaling of the magnetic spring stiffness with $d_{Stator}$ in Fig. 2b, we derive (in the supplemental material S2) a proportionality relation (of constant $\alpha$) between $T_{amp}$ and a function $g(d_{Stator})$: $T_{amp} = \alpha \times g(d_{Stator})s$. This function $g(d_{Stator})$ treats the rotor as an equivalent magnetic dipole moment ($\vec{m}_R$), but, computes $B_S$ based on the manufacturer's models (cf. supplemental material S2) [25] [26] [27]. The theoretical estimation finds a proportionality constant $\alpha = 7.57$ N m.

Alternatively, fitting the experimental points in Fig. 2b of the nonlinear (sinusoidal) spring stiffnesses to the derived relation results in an $\alpha_{Fit} = 7.16$ N m which is within ~6% of the theoretical estimation. We note in Fig. 2b that this model captures the experimental scaling between magnetic spring stiffness and the stator-to-rotor distance. Hence, we conclude that the derived model (cf. supplemental material S2) is suitable for estimating stators magnetic field and restoring torque for the considered stator-to-rotor distances. However, in general, the magnetic dipole moment approximation of the rotor does not lead to such acceptable estimations of magnetic torque and numerical simulations are required [28].

In addition to the fit of the derived model, we fit the experimental data in Fig. 2b to the following power model:

$$y = A \times d_{Stator}{}^n \tag{3}$$

where $y$ in this case corresponds to $K_{Mag}$. In fact, we apply the function in (3) as an empirical fit to quantify the power low (of pre-factor $A$ and exponent $n$) relating any measured parameter $y$ in our study to the stator-to-rotor distance $d_{Stator}$. As for linearized-magnetic-torsional stiffness, the power fit of Fig. 2a demonstrates that the experimental values of $K_{Mag}$ scale with almost $d_{Stator}^{-2}$.

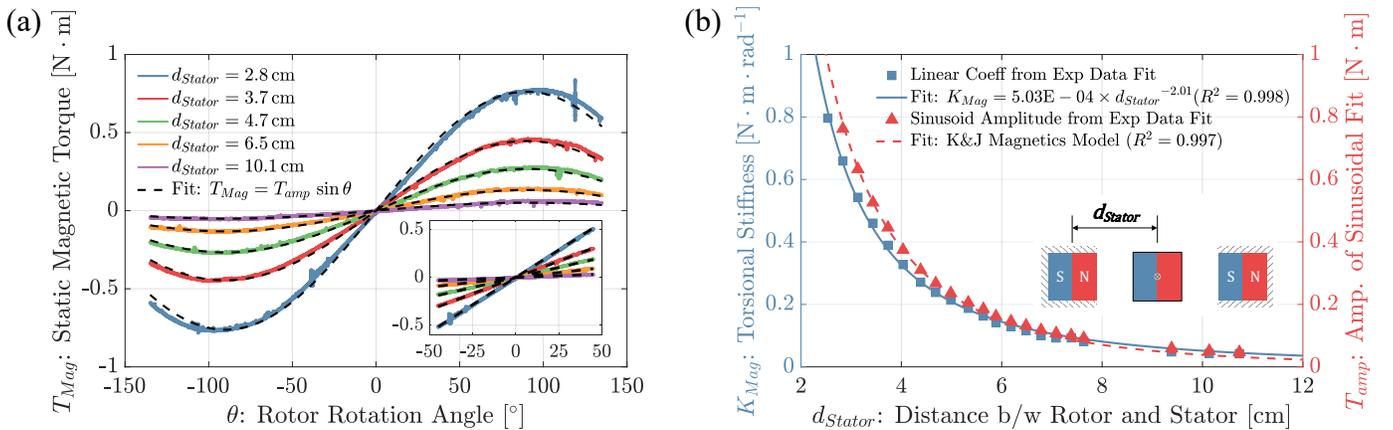

**Fig. 2.** Torsional magnetic spring static response. (a) Measured magnetic restoring torque, $T_{Mag}$, exerted by the PM stators on the PM rotor as function of the rotor's angular rotation $\theta$ w.r.t. its equilibrium position for different stator-to-rotor distances $d_{Stator}$. Inset shows small amplitude (linearized) response of $T_{Mag}$ for $\theta \in [-45°, +45°]$. Experimental data (colored lines) are fitted (dashed lines) to (1) for $\theta \in [-135°, +135°]$ and to (1) for $\theta \in [-45°, +45°]$ (inset). (b) Data points showing the dependence of the linear $K_{Mag}$ (small angle fit coefficient, left blue axis) and the nonlinear $T_{amp}$ (sinusoidal stiffness coefficient, right red axis) on $d_{Stator}$. The fitting line use the relations described in the legend for small angle approximation (blue-solid line, left axis) and large sinusoidal approximation (red-dashed line, right axis). The range of the 95% confidence intervals on every stiffness parameter identified in fitting is less than 0.2% of the parameter value.

## 4. Ringdown Response of The Mechanical System

In order to study the dissipations in this magnetic device, we investigate the ringdown (free) response of the rotor after releasing it from an initial angular displacement. During this ringdown response, the rotor undergoes decaying oscillations due to the magnetic stiffness mechanism of section III and the dissipations induced by the ball bearings, the magnets and the surrounding air. This section focuses on modeling and identifying the dissipation mechanisms in this magnetic device as function of the stator-to-rotor distance $d_{Stator}$.

*Reduced-Order Model (ROM) of the Ringdown Response*

The damping in the mechanical oscillations arises mainly from the ball bearings and the core losses in the permanent magnets [24]. The core losses include any form of dissipation related to the magnetic field variation like eddy currents and magnetic material hysteresis losses [24]. In this work, the collective effect of all dissipations on the system's dynamics is modeled using two types of damping: solely-viscous damping and viscous plus dry (combined) damping.

Viscous damping results in dissipation torque $T_{Vis} = -c\dot{\theta}$ proportional to the rotor angular velocity $\dot{\theta}$ with $c$ being the viscous damping coefficient in $N \cdot m \cdot s \cdot rad^{-1}$ [29]. The combined damping model considers a dry (Coulomb) torque $T_{Dry}$ in addition to the viscous damping torque $T_{Vis}$ [29]. However, the Coulomb damping $T_{Dry}$ is characterized by a constant amplitude equal to the Coulomb friction coefficient $T_f$ (in $N \cdot m$) that acts against the motion of the rotor according to $T_{Dry} = -T_f \operatorname{sgn} \dot{\theta}$ with the signum function defined by [29]:

$$\operatorname{sgn} \dot{\theta} = \begin{cases} +1 \text{ for } \dot{\theta} > 0 \\ \phantom{+}0 \text{ for } \dot{\theta} = 0 \\ -1 \text{ for } \dot{\theta} < 0 \end{cases}. \tag{4}$$

For ball bearings, the Coulomb friction coefficient $T_f$ is typically proportional to some bearing's characteristic radius and to the radial pressure supported by that bearing (designated by $P$ in the inset of Fig. 3b) [19]. During ringdown of the proposed torsional magnetic oscillator, combined damping results the equation of motion in (5):

$$J\ddot{\theta} + c\dot{\theta} + K_{Mag}\theta + T_f \operatorname{sgn} \dot{\theta} = 0 \tag{5}$$

with $J$ denoting the moment of inertia about the axis of rotation of the rotor assembly of Fig. 1b.

Fig. 3 and (5) assume linear stiffness $K_{Mag}$ of the magnetic spring. This linear stiffness assumption is adopted because the angular amplitudes in the experiments are less than 15° (Fig. 3c-d). The equation of motion in (5) is typically expressed in terms of normalized parameters:

$$\ddot{\theta} + 2\zeta\omega_n\dot{\theta} + \omega_n^2\theta + \omega_n^2\theta_f \, \text{sgn}\, \dot{\theta} = 0 \tag{6}$$

In (6), $\omega_n$, $\zeta$, and $\theta_f$ denote the natural frequency of the system, the viscous damping ratio and the dry friction angle. Each of these normalized parameters describes a system's characteristic observable in the angular free response.

The natural frequency $\omega_n = 2\pi f_n = 2\pi/T_n = \sqrt{K_{Mag}/J}$ is the angular frequency (in $\text{rad} \cdot \text{s}^{-1}$) of the system's oscillations in the absence of all forms of damping; where $f_n$ and $T_n$ are the oscillations' natural frequency (in Hz) and period (in s), respectively. For viscously under-damped oscillations, the frequency of oscillations reduces from $\omega_n$ to the damped frequency $\omega_d = \omega_n\sqrt{1-\zeta^2}$ where the viscous damping ratio $\zeta = c/(2\sqrt{J\,K_{Mag}}) < 1$. Specifically, $\zeta$ dictates the logarithmic decrement in amplitude of oscillations (cf. Fig. 3a) such that:

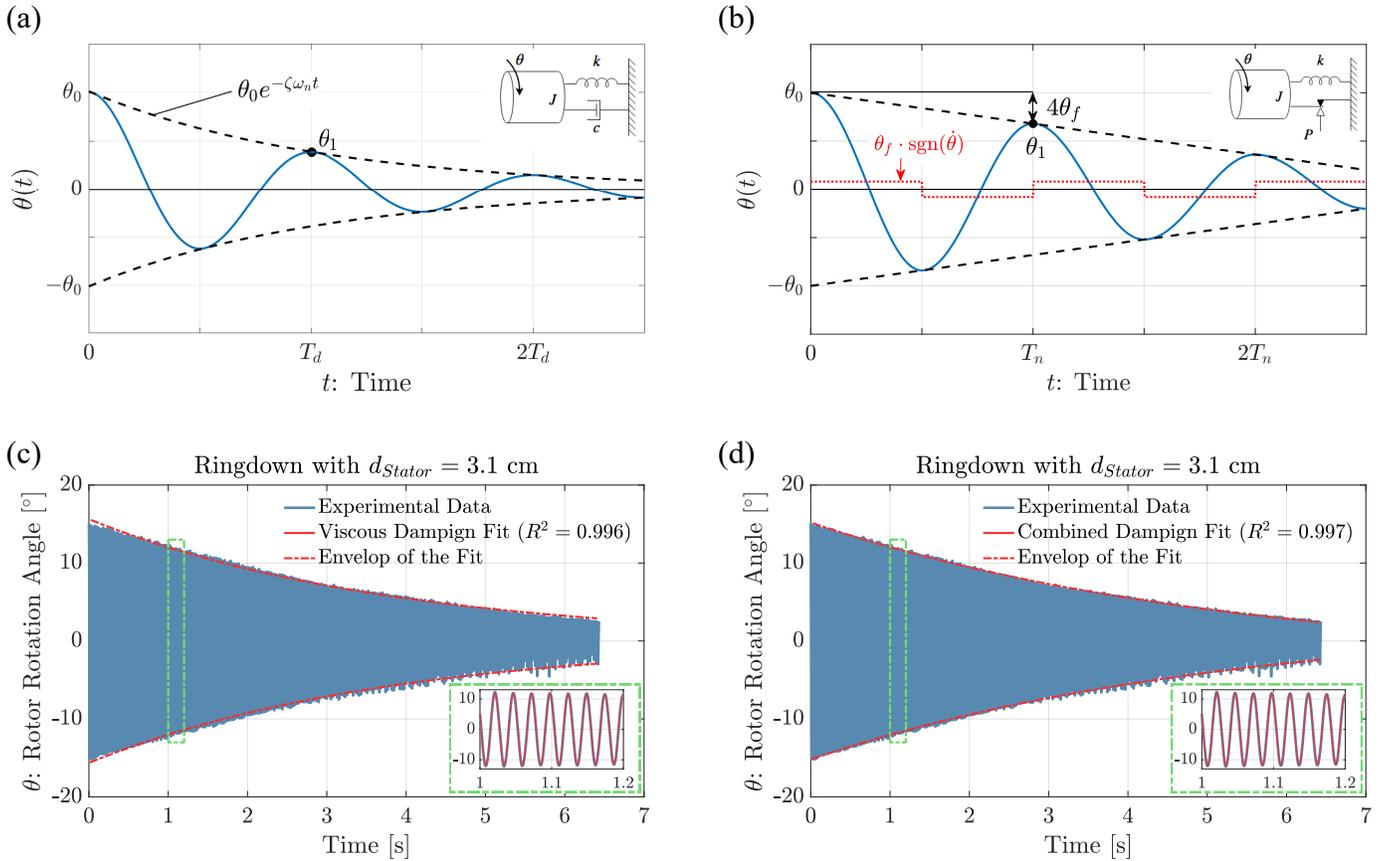

**Fig. 3.** Magnetic oscillator ringdown (free) response. Simulation of the decay of rotor oscillations, after being released from an intial angular displacement $\theta_0$, under the effect of (a) solely-viscous damping or (b) dry friction. Measured rotor's angular displacement $\theta$ as function of time in ringdown experiment with $d_{Stator} = 3.1$ cm (best $R^2$ among all measurements) in addtition to the decay envelop of the system resulting from fitting (c) viscous damping model and (d) combined damping model. The insets of (c) and (d) show a zoomed view of measurements between 1 s and 1.2 s (inside green box) in addition to the instantaneous time fit of the respective damping model.

$$\theta(t) = \Theta_0 e^{-\zeta\omega_n t} \cos(\omega_d t + \phi_0) \tag{7}$$

where $\Theta_0$ and $\phi_0$ denote the undamped amplitude and phase shift of oscillations, respectively.

On the other hand, under the action of dry (Coulomb) friction alone (Fig. 3b), the rotor oscillates with a period equal to the natural period $T_n = 2\pi/\omega_n$ during which the oscillations' amplitude decreases by a fixed amount of $4\theta_f$ – where the dry friction angle $\theta_f = T_f/K_{Mag}$ – as depicted by the linear envelope of Fig. 3b.

However, combined (viscous and dry) damping model in (6) modulates the oscillations with a combination of logarithmic and linear decrement envelopes according to (cf. supplemental material section S3):

$$\theta_k(t) = \left[\Theta_0 - 2\theta_f \frac{\exp(k\pi\zeta/\sqrt{1-\zeta^2})-1}{\exp(\pi\zeta/\sqrt{1-\zeta^2})-1} \cdot \frac{e^{\zeta\omega_n t_1}}{\sqrt{1-\zeta^2}}\right] e^{-\zeta\omega_n t} \cos(\omega_d t + \phi_0) - (-1)^k S_0 \theta_f \tag{8}$$

where $\theta_k$ describes the rotor's angle during between the $k^{\text{th}}$ and the $(k+1)^{\text{th}}$ extrema of oscillations ($k \in \mathbb{N}$). In (8), $t_1$ and $S_0$ denote the instant of achieving the first extremum of oscillations (expressed by (s22) in S3) and the sign of initial angular velocity (expressed by (s28) in S3). For full derivation of (8), we refer the reader to the supplemental material section S3.

*Experimental Validation*

In order to experimentally validate the reduced order model in (5), we investigate the ringdown response by releasing the rotor from an initial angle for varying stator-to-rotor distance as described in the supplemental material S1. To approximate the magnetic spring with the linearized stiffness of (5), we consider the time segments with low oscillations' amplitude (i.e., 15° and 2.5° as shown in Fig. 3c-d).

For every stator-to-rotor distance, we postprocess and fit the low-amplitude-ringdown response to both damping models (cf. supplemental material S3). Fitting to combined damping model in (8) results in $\{\omega_n, \zeta, \theta_f\}$; whereas viscous damping model in (7) results in $\{\omega_n, \zeta\}$ (in addition to $\Theta_0$ and $\phi_0$). We examine the results of both fitting forms in order to unravel the importance of incorporating the dry friction effect on modeling the rotor's free response.

For example, Fig. 3c and 3d show the same experimental ringdown response fitted to the two different models as explained earlier. These figures and their insets depict that both fits track very well the experimental angular displacement ($R^2 > 0.995$). Thus, given their good fitting performance, the identified parameters for different $d_{Stator}$ characterize experimentally the magnetic effect on the (viscous and dry) dissipations as depicted by Fig. 4.

Regarding ringdown frequency, Fig. 4a displays the values of natural frequency $f_n = \omega_n/(2\pi)$ calculated from fitting with the combined and solely-viscous damping models. The two models result in almost the same natural frequencies that increase almost inversely proportional with the stator-to-rotor distance as indicated by the empirical fit of (3) on Fig. 4a.

Remarkably, the linear magnetic stiffnesses ($K_{Mag}$) of Fig. 2b scales linearly with the square of the natural angular frequencies ($\omega_n^2$) deduced from identified frequencies of Fig. 4a ($f_n$) as depicted by the inset of Fig. 4a. This experimental linearity represents the fact that $K_{Mag} = J(2\pi f_n)^2 = J\omega_n^2$ allowing to deduce the experimental $J = 8.99 \times 10^{-6}$ kg·m². This experimental value of the moment of inertia agrees very well (within 90%) with its value ($J = 1.01 \times 10^{-5}$ kg·m²) estimated using SolidWorks CAD software and the mass properties of Neodymium [27] and titanium (from SolidWorks library).

As for viscous damping effects, Fig. 4b shows the viscous damping ratios $\zeta$ and coefficients $= 2J\omega_n\zeta$ identified from fitting to the two models. The combined damping model in Fig. 4b presents increasing viscous damping ratios ($\zeta$) but decreasing viscous damping coefficients ($c \sim d_{Startor}^{-0.47}$) with increasing $d_{Stator}$. Conversely to the monotonic trends of the combined damping parameters, the viscous damping ratios ($\zeta$) and coefficients ($c$) of the solely-viscous damping model illustrate an intriguing local minimum around $d_{Startor} = 3.7$ cm depicted in Fig. 4b. Notably, the values of $c$ from the solely-viscous damping model experience a less steep increase on the right of the local minimum than on its left where $c$ grows as

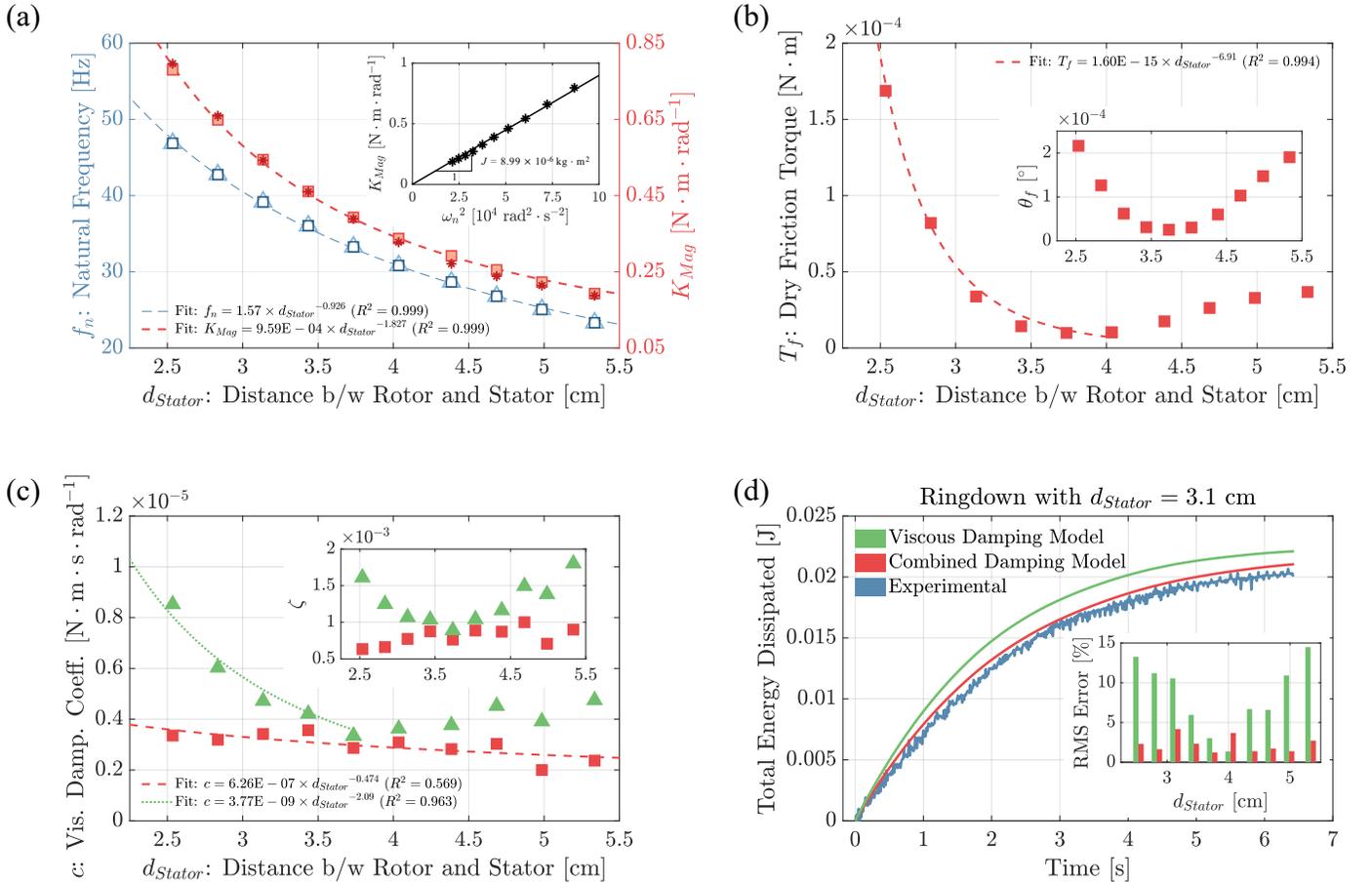

**Fig. 4.** Effect of $d_{Stator}$ on: (a) the natural frequency $f_n$ (left blue axis) and the linear magnetic stiffness $K_{Mag}$ (right red axis), (b) the viscous damping coefficient $c$ and damping ratio $\zeta$ (inset), (c) the dry friction torque $T_f$ and friction angle $\theta_f$ (inset). Plotted points correspond to the parameters identified from static (asterisks) and ringdown measurements by fitting to solely-viscous (triangles) and combined (squares) damping models. The viscous and combined damping datapoints are fitted to the dotted and dashed curves, respectively. Inset of (a) displays $K_{Mag}$ from static measurements vs. $\omega_n^2$ from ringdowns for the same $d_{Stator}$; along with their linear fit (solid line) whose slope equals the experimental moment of inertia of the rotor $J$. The range of the 95% confidence intervals with respect to identified value are less than 0.002% for $K_{Mag}$, and less than 1.5% for $c$ and $T_f$. (d) Temporal evolution of energy dissipated in the system with $d_{Stator} = 3.1$ cm (best R²) as found experimentally (blue) and as estimated in the combined (red) and the solely-viscous (green) damping models. Inset shows the relative root-mean-square (RMS) error in the evolution of the energy dissipation as estimated in the combined (red) and the solely-viscous (green) damping models during the ringdown measurements for each $d_{Stator}$.

$d_{Startor}^{-2.1}$ (Fig. 4b).

Physically though, the viscous damping strengthens with the increase of magnetic related losses (i.e., eddy currents and core losses) which typically increase with the increase in frequency and amplitude of the oscillating magnetic field [24]. Since both the rotor frequency (Fig. 4a) and the amplitude of the magnetic field between the rotor at the stators decrease with increasing $d_{Startor}$, the viscous damping coefficient ($c$) weakens with increasing $d_{Startor}$ as observed by the values identified in combined damping model. Hence, we conclude that the solely-viscous damping does not agree with the mechanisms of the magnetic losses [24].

As for the Coulomb damping, the combined damping model exhibits in Fig. 4c a local minimum in dry friction angles ($\theta_f$) and the dry friction torques ($T_f = J\omega_n^2\theta_f$) around $d_{Stator} = 4.04$ cm. Reducing stator-to-rotor distance below 4.04 cm sharply increases friction torque $T_f$ at a rate proportional to $d_{Startor}^{-6.9}$. Physically, the increase in $T_f$ mainly increases due to radial pressure ($P$ in the inset of Fig. 3b) on the rotor's bearings (cf. Fig. 1b) resulting from the net magnetic force exerted by the stators [20] [29]. Ideally, the rotor's equidistant location from the two stators provides a symmetry that prevents the creation of net transverse magnetic pressure. In practice however, this symmetry breaks due to unavoidable misalignments leading to a net transverse magnetic pressure that minimizes for $d_{Stator} \approx 4.04$ cm as discovered experimentally.

*Comparison between the Damping Models*

For a quantitative comparison between the two damping models (viscous and combined), we investigate the error in predicted energy dissipation (cf. Fig. 4d and supplemental material section S4). As an experimental reference for dissipated energy, we consider the lost mechanical energy (kinetic and elastic) at the instant ($t$) using the identified $J$ and $K_{Mag}$ of Fig. 4a. We compare this experimental reference of the dissipated energy with the dissipative work exerted according to the combined and the solely-viscous damping models (cf. supplemental material S4). Fig. 4d shows an example of the computed dissipated energies as function of time for $d_{Stator} = 3.1$ cm. We see that the combined damping model tracks the change in energy dissipation with time more accurately than the solely-viscous damping model.

To quantitatively evaluate this performance the models for every $d_{Stator}$, the inset of Fig. 4d displays the relative root-mean-square (RMS) that computes the error between the experimental energy dissipation and the dissipation estimated by each model (cf. supplemental material S4). This inset indicates that the combined damping model predicts the time evolution of energy dissipation more accurately for most of the stator-to-rotor distances: the relative error <5% in combined damping model but >13% in the solely-viscous damping model for some values of $d_{Stator}$.

For instance, the inset of Fig. 4d shows that the error in combined damping model distributes randomly with respect to $d_{Stator}$, whereas the error in the solely-viscous damping model increases as $d_{Stator}$ deviates from 4.04 cm. Recall that near this same stator-to-rotor distance of 4.04 cm, the dry friction torque $T_f$ achieves a minimum in Fig. 4c. Hence, the improved performance of the solely-viscous damping model results from the reduction of dry friction in the actual system. Therefore, the combined damping model eliminates the observed error dependence on $d_{Stator}$ in the solely-viscous damping model by accounting for the different damping mechanisms involved during the ringdown response.

## 5. Electromechanical Coupling

To proceed from the ringdown dynamics of section IV towards forced response dynamics, we force the rotor's oscillation by a nearby electromagnetic coil (cf. Fig. 1). Therefore, in this section, we model the interaction between the rotor and the coil by accounting for the electromechanical coupling introduced by the coil, namely: the torque ($T_{Coil}$) exerted by the coil magnetic field and the back-electromotive force ($u_{EMF}$) induced across the coil due to rotor rotation as depicted by Fig. 5a,c respectively.

*Torque Coupling Coefficient*

In order to develop reduced order models of the electromechanical coupling, we approximate the rotor by a magnetic dipole moment (cf. section III). This dipole approximation expresses the driving torque ($T_{Coil}$) applied on the rotor of angle $\theta$ due to the flow of electric current ($i_{Coil}$) thorough the coil according to the following nonlinear relationship (cf. supplemental material section S5):

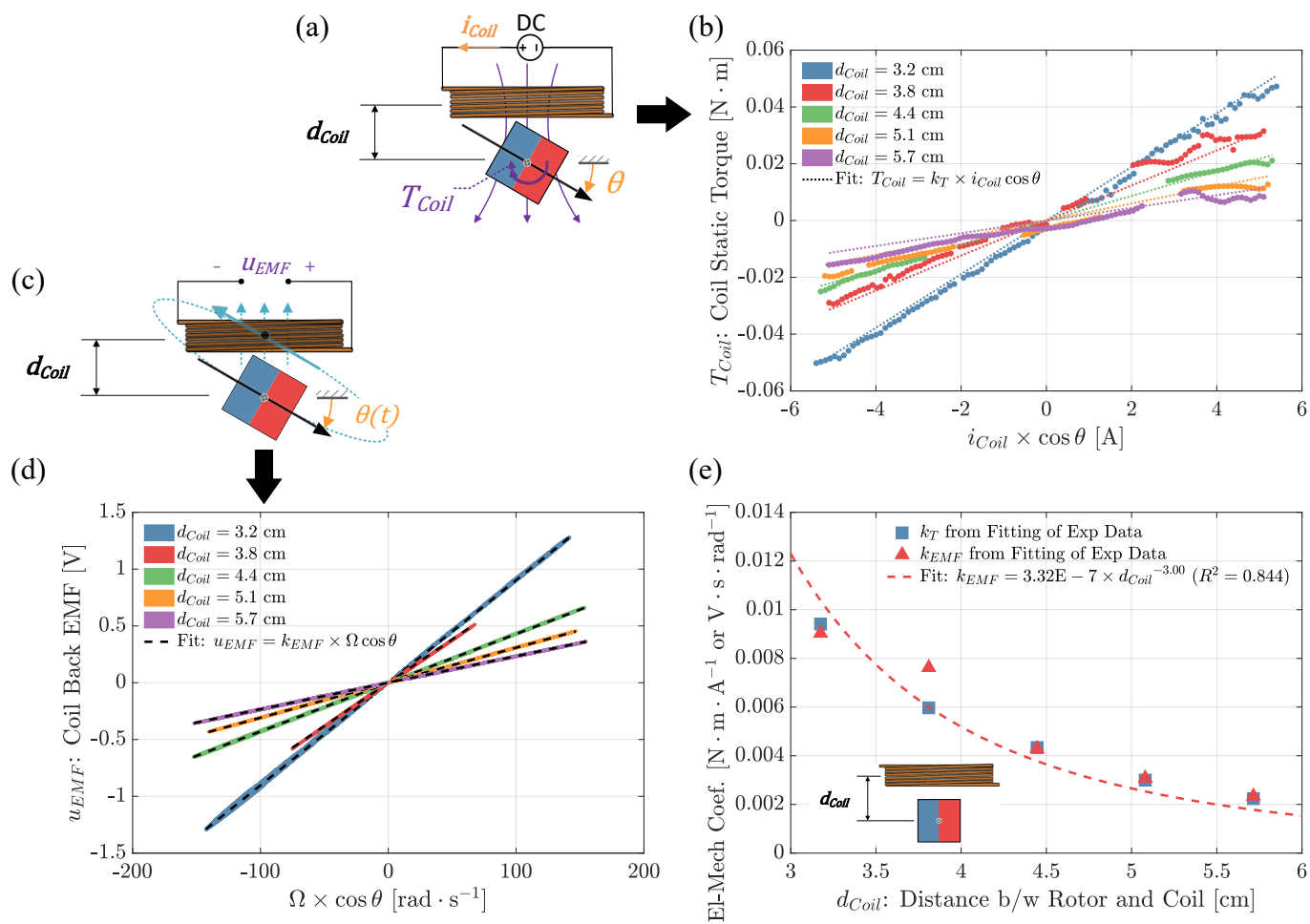

**Fig. 5.** Electromechanical coupling. (a) Schematic depicting the experiment conducted to measure the torque $T_{Coil}$ applied by the coil (subjected to a DC voltage) while quasi-statically rotating the rotor. (b) Measured $T_{Coil}$ as function of the DC current flowing throw the coil $i_{Coil}$ and the rotor angular position $\theta$ for different coil-to-rotor distances $d_{Coil}$. (c) Schematic depicting the experiment conducted to measure back electromotive force $u_{EMF}$ induced across the coil while the rotor fully rotates with angular speed $\Omega$. (d) Measured $u_{EMF}$ as function of rotor angular speed $\Omega$ and position $\theta$ for different $d_{Coil}$. (e) Effect of $d_{Coil}$ on electromechanical coupling coefficients $k_T$ and $k_{EMF}$. The range of the 95% confidence intervals on $k_T$ and $k_{EMF}$ is less than 0.35% of each identified value.

$$T_{Coil} = k_T \times i_{Coil} \times \cos\theta. \tag{9}$$

To identify experimentally the torque coupling coefficient $k_T$ of (9), we apply a dc current (via a BK Precision 1693 high power supply) into the coil with the absence of stators as shown in Fig. 5a (cf. supplemental material S5). Fig. 5b presents the measured values of the quasi-static $T_{Coil}$ against the product of $i_{Coil} \times \cos\theta$ for five different coil-to-rotor distances $d_{Coil}$. For every coil-to-rotor distance, the experimental points of Fig 6a demonstrate the linear relation between $T_{Coil}$ and $i_{Coil} \times \cos\theta$ as intended by (9). Fitting these measurements to the model in (9) results in the values of $k_T$ plotted versus their respective $d_{Coil}$ as shown in Fig. 5e. The torque coupling coefficient decreases with the increased coil-to-rotor distance which can be explained by the decrease in coil magnetic field at the rotor location.

*Back-Electromotive Force Coupling Coefficient*

Alternatively, the back-electromotive force $u_{EMF}$ induced across the coil results from the change in its magnetic flux due to the nearby rotor's PM movement as illustrated by Fig. 5c. In order to model $u_{EMF}$, we also approximate the rotor by its magnetic dipole moment, which leads to the following nonlinear expression (cf. supplemental material S5):

$$u_{EMF} = k_{EMF} \times \Omega \times \cos\theta \tag{10}$$

where $\Omega = \frac{d\theta}{dt}$ denotes the rotor angular velocity.

Experimentally, we identify the back-EMF coupling coefficient $k_{EMF}$ in (10) by measuring the voltage drop $u_{EMF}$ across the open-loop coil as function of the rotor full rotations in the absence of the stators (cf. Fig. 5c and supplemental material S5). Fig. 5d presents the experimental $u_{EMF}$ versus $\Omega \times \cos\theta$ for every $d_{Coil}$ measured where the data clearly exhibits the linear dependence in (10). Hence, we fit the measurements on Fig. 5d to the model in (10) to identify the experimental $k_{EMF}$ for every $d_{Coil}$. Fig. 5e groups all these experimental $k_{EMF}$ whose values decrease with increasing values of coil-to-rotor distances.

For instance, the dipole approximation of the rotor results in a dependence of the form $k_{EMF} = \beta d_{Coil}^{-3}$ with $\beta \approx 2.90 \times 10^{-7}$ V·s·m³·rad⁻¹ in our system (cf. supplemental material S5). However, the fitting of the experimental $k_{EMF}$ of Fig. 5e to the physical model of $k_{EMF} = \beta_{Fit} d_{Coil}^{-3}$ leads to $\beta_{Fit} = 3.32 \times 10^{-7}$ V·s·m³·rad⁻¹ which is within 15% of the value estimated by the model. Therefore, the model in (10) reflects the physics of the back EMF coupling with a reasonable deviation from the experimental results compensating for the assumed dipole approximation of the rotor.

Moreover, Fig. 5e indicates an equality between the two experimental electromechanical coupling coefficients ($k_T$ and $k_{EMF}$) for all but one coil-to-rotor distances. Actually, this equality demonstrates a conservative (lossless) electrical and mechanical conversion through the magnetic coupling, which is generally observed in most electric machinery [27] (cf. supplemental material S5).

## 6. Frequency Sweep

In this section, we investigate the forced frequency response of the magnetic system by adopting the nonlinear models of sections III, IV, and V. For this purpose, the coil studied in section V drives the rotor into frequency sweeps as shown in Fig. 6 and 7 (see supplemental video 2 where the rotor is forced around resonance). In these forced responses, the models of the previous sections are lumped into the electromechanical circuit representation of Fig. 6a.

In this circuit, the AC voltage source corresponds to the (measured) voltage applied by the power amplifier across the terminal of the coil. The coil itself accounts in Fig. 6a for: the electrical inductance $L$ (due to its self-inductance), the electrical resistance $R$, and the non-linear gyrator [30]. This gyrator relates the electric current (\rotational velocity) in the electrical (\mechanical) system to the magnetic torque (\back-EMF) in the mechanical(\electric) system [30] based on the electromechanical coupling model of section V.

Moreover, in the circuit of Fig. 6a, equivalent electric components model the mechanical system (rotor and stators): the electric inductance $J$ representing the rotor inertia, the nonlinear capacitor representing the nonlinear magnetic stiffness, the electric resistance $c$ representing the viscous damping, and the DC voltage source connected in the rectifier to always oppose the voltage supply representing the Coulomb friction [29]. This circuit of Fig. 6a leads to the following system of equations:

$$\begin{cases} L\frac{di_{Coil}}{dt} + Ri_{Coil} + k_{EMF}\dot{\theta}\cos\theta = U_{in}(t) \\ J\ddot{\theta} + c\dot{\theta} + T_f\,\text{sgn}\,\dot{\theta} + T_{amp}\sin\theta = k_T i_{Coil}\cos\theta \end{cases}. \quad (11)$$

To validate the model in (11), we compare the simulated to the experimental frequency sweeps in Fig. 6b (forward sweep) and 7c (backward sweep) for the device with $d_{Stator} = 3.4$ cm and $d_{Coil} = 3.2$ cm (cf. the supplemental material section S6 for detailed description of the experiments and the simulations).

The parameters used to run the simulation are listed in Table I. The table values are identical to the identified values of $J$, $c$, and $T_f$ in section IV (using the combined damping model) and of $k_T$ in section V for $d_{Stator} = 3.4$ cm and $d_{Coil} = 3.2$ cm. However, the stiffness coefficient in Table I is calculated using $T_{amp} = J(2\pi f_n)^2$ where $f_n$ is the natural frequency of the system when the nonlinearity is not excited for

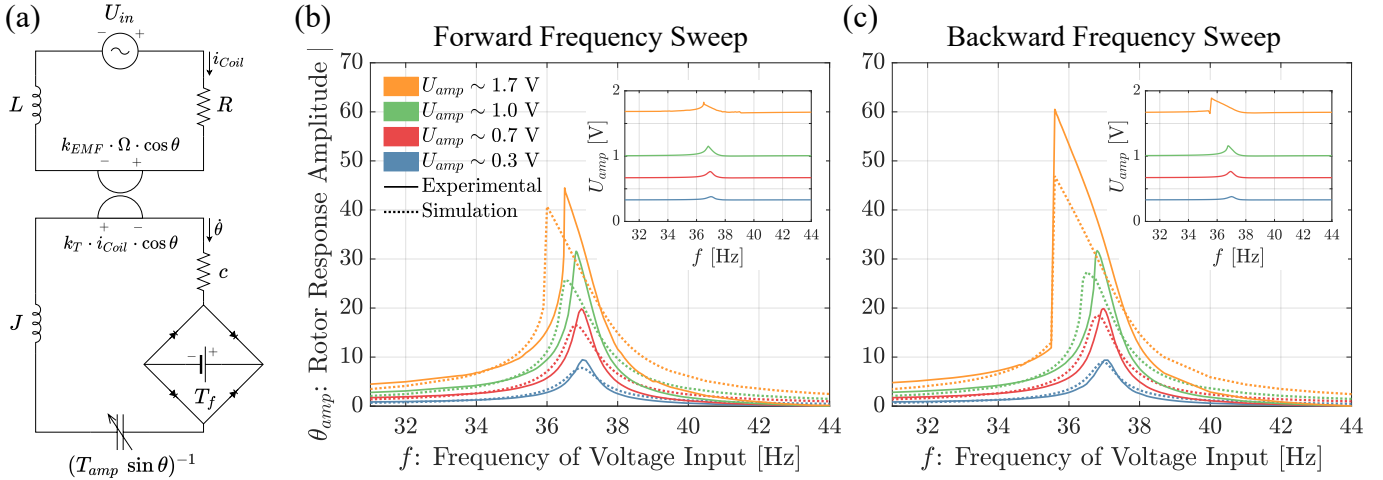

Fig. 6. Frequency sweep response of the rotor. (a) Equivalent circuit representation of the electromechanical system governing the dynamics of the rotor actuated by the coil. For the system with $d_{Stator} = 3.4$ cm and $d_{Coil} = 3.2$ cm, experimental (solid lines) and simulation (dotted lines) angular amplitude $\theta_{amp}$ of the steady state rotor response in (b) forward/increasing and (c) backward/decreasing sweeps in frequency $f$ of the driving harmonic voltage input $U_{in} = U_{amp}\cos(2\pi ft)$ at different driving levels. Insets in (b) and (c) display the amplitude $U_{amp}$ at the respective frequency $f$ and driving level.

14**Table I**
Values for parameters used to simulate the system (11)

| $R = 1.76 \, \Omega$ | $k_{EMF} = 9.41 \, \text{mV} \cdot \text{s} \cdot \text{rad}^{-1}$ | $J = 8.99 \times 10^{-6} \, \text{kg} \cdot \text{m}^2$ | $c = 3.17 \times 10^{-6} \, \text{N} \cdot \text{m} \cdot \text{s} \cdot \text{rad}^{-1}$ |
|---|---|---|---|
| $L = 1.83 \, \text{mH}$ | $k_T = 9.41 \, \text{mN} \cdot \text{m} \cdot \text{A}^{-1}$ | $T_{amp} = 0.484 \, \text{N} \cdot \text{m}$ | $T_f = 8.54 \times 10^{-5} \, \text{N} \cdot \text{m}$ |

weak forcing level (i.e., $U_{amp} \sim 0.3$ V in Fig. 6b-c). This $T_{amp}$ of Table I does not greatly diverge from the one identified in static measurements of section III ($< 5\%$ difference); but we consider this correction in order to correct for the variations in the setup that might have occurred between the measurement dates of the current and the previous sections. Finally, we identify the coil self-inductance $L$ and resistance $R$ in Table I by running frequency sweeps for the coil as explained in the supplemental material section S1.

In the conducted frequency response, we drive the system with a harmonic voltage input $U_{in}(t) = U_{amp} \cos(2\pi f t)$ by increasing\decreasing $f$ in the forward\backward sweep as shown in Fig. 6b-c and Fig. 7. To capture the nonlinear hysteresis between the forward and the backward sweeps (cf. Fig. 7a for a clearer illustration), the implemented numerical method accounts for the state continuity in the system's dynamics during the sweeps. In this numerical method, the coil voltage $U_{in}(t)$ in (11) is computed from the experimental voltage amplitudes $U_{amp}$ measured using lock-in amplifier for every frequency of excitation $f_{ext}$ as recorded in the insets of Fig. 6b,c (cf. supplemental material section S1.D for more info about the experimental measurements of $U_{amp}$ and $f_{ext}$. In effect, for every applied $f_{ext}^k$ of index $k$, we compute the respective instantaneous voltage input $U_{in}^k(t)$ in (11) using:

$$U_{in}^k(t) = U_{amp}^k \cos(2\pi f_{ext}^k \cdot t). \tag{12}$$

With the objective to capture the forward (Fig. 6b) and backward (Fig. 6c) sweeps, we impose the system to initial conditions of $\theta$ and $\dot{\theta}$ that mimic the effect of the sweep direction on the nonlinear response of the rotor:

1) We start by simulating the system with trivial (zero) initial conditions of $\theta$ and $\dot{\theta}$ by applying the first $U_{in}^0(t)$ for a duration sufficient to reach steady-state.
2) Then, we apply a Fast-Fourier-Transform (FFT) of to find the amplitude $\theta_{amp}^0$ of the steady-state cycles of frequency $f_{ext}^0$.
3) For the next excitation frequency, the system is simulated from the following set of initial conditions with $k = 1$:

$$\begin{cases} \theta^k(t=0) = \theta_{amp}^{k-1} \\ \dot{\theta}(t=0) = 0 \end{cases} \tag{13}$$

By this same procedure (steps 2 and 3) at every successive step $k$, the simulation plots in Fig. 6b,c are generated where we plot the $\theta_{amp}^k$ that results from the FFT (in step 2) at every $f_{ext}^k$.

In fact, this numerical method enforces the simulations to demonstrate the nonlinear hysteresis loops observed in Fig. 6b-c where softening nonlinearity presents a backward sweep that occurs with larger amplitude and at a lower jump-frequency than the forward sweep [31]. Hence, the developed models in the previous sections capture the type of nonlinearity in the device with $d_{Stator} = 3.4$ cm and $d_{Coil} = 3.2$ cm.

Moreover, for low voltage amplitudes ($U_{amp} \sim 0.3$ V and $\sim 0.7$ V in Fig. 6b and 7c), these models acceptably predict the magnitude of the experimental frequency sweeps.

However, for large amplitudes ($U_{amp} \sim 1.0$ V and $\sim 1.7$ V in Fig. 6b and 7c), the simulations greatly deviate from the measurements, which might not only result from the models' deficiency in estimating the response at large amplitudes, but also from the inaccuracy in the simulation parameters in Table I. These parameters (except for $K_{Mag}$ as explained earlier) are assigned the same values identified in the isolated experiments of the previous sections. Thus, they do not account for modification introduced experimentally by the different setup assemblies performed during this study. Actually, Fig. 6b and 7c shows that damping identified from the ringdown experiments of section IV overestimates the actual damping in the frequency sweep experiments.

*Additional Observed Nonlinearities*

Although Fig. 6b-c examines the frequency sweep response for $d_{Stator} = 3.4$ cm, we are interested in the frequency weep response and the nonlinearity for even smaller $d_{Stator}$. Therefore, we present in Fig. 7a-d the frequency sweeps for $d_{Stator} \in \{3.4, 3.1, 2.8, 2.5 \text{ cm}\}$, respectively. In Fig. 7a-d, a (green) line joins the maximum $\theta_{amp}$ attained in the sweep of the particular $d_{Stator}$. This line represents the experimental backbone curve of the system for the respective $d_{Stator}$, which characterizes the type of nonlinearity [31].

For instance, $d_{Stator} = 3.4$ and 3.1 cm (Fig. 7a-b, respectively) depict a softening nonlinearity where the frequency decreases with amplitudes along the backbone curves. However, the two smallest $d_{Stator} = 2.8$ and 2.5 cm (Fig. 7c-d) demonstrate a stiffening behavior at the small forcing levels, then a softening behavior at the larger forcing levels. Actually, this observed stiffening-to-softening nonlinearity cannot be modeled by the spring model in (1) which only exhibits a softening nonlinearity by the sinusoidal dependence between $T_{Mag}$ and $\theta$.

Hence, (1) is not appropriate to model the torsional magnetic spring at small stator-to-rotor distances (for $d_{Stator} \leq 2.8$ cm). This drift in nonlinearity could be anticipated from the static measurements of Fig. 3a.

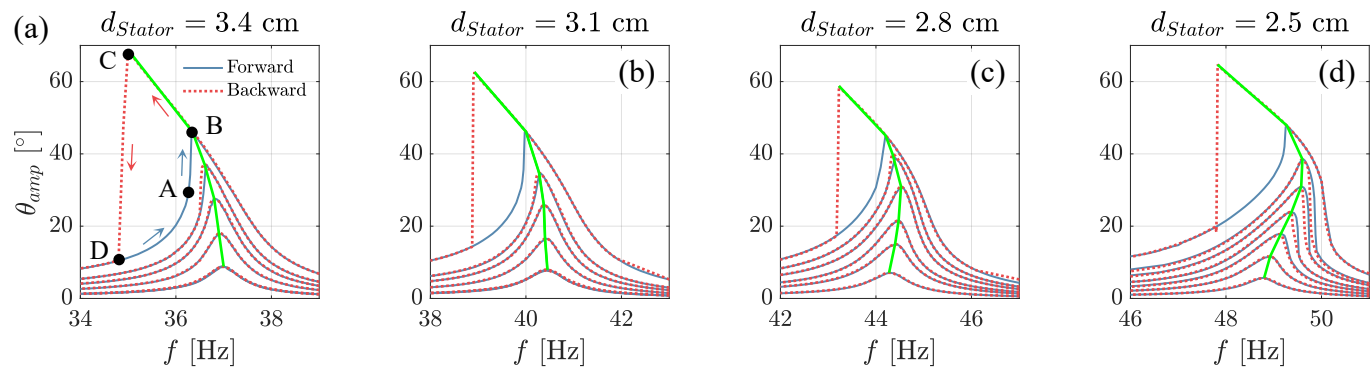

**Fig. 7.** Effect of stator-to-rotor distance on the system nonlinearity. Experimental measurements of angular amplitude $\theta_{amp}$ of the steady state rotor response in forward (blue-solid) and backward (red-dotted) frequency sweeps for increasing amplitudes in driving voltage with $d_{Stator}$ of (a) 3.4 cm, (b) 3.1 cm, (c) 2.8 cm, and (d) 2.5 cm. The green line corresponds to the backbone curve (up to the measurements' resolution) that joins the points of maximum $\theta_{amp}$ in every frequency sweep. In Fig. 7a, the loop A-B-C-D-A designates the nonlinear hysteresis where the single frequency steps A→B and C→D are the nonlinear discontinuity jumps in forward and backward sweeps, respectively.

This figure shows that the model in (1) capture less properly the torque-angle dependence for $d_{Stator} = 2.8$ cm. Therefore, the stiffness model needs an update in order to capture the magnetic spring effect for reduced distances between magnets.

## 7. Conclusion

In this work, we present a design a torsional-magnetic spring oscillator while examining its performance and modeling. We measure the different characteristics of the system including torques, angular displacements, and damping parameters. We experimentally validate the coupled dynamics of the identified mechanical and electric based on their performance in the frequency sweep response. Finally, the study investigates the effects of stator-to-rotor and coil-to-rotor distances by the mean of experimental parametric sweeps.

The study demonstrates the limits on the use of the magnetic-dipole-moment-approximation for small rotor-to-stator distances. This magnetic-dipole-moment-approximation of the rotor along with approximating the coil with an inductance-equivalent current loop located at the coil center presents acceptable modelling results of the coil electromechanical effect for all the coil-to-rotor distances examined.

The work emphasizes on the importance of considering the dry friction exerted by the bearings in the rotor's free response (ringdown). For this part of the study, we derive an analytical expression that explicitly express the time evolution of angular displacement under the effect of combined (viscous and dry) damping as function of general ringdown initial conditions.s This explicit relation provides a computational advantage for the system identification as compared to simulating the system's equation of motion. The ringdown identifications at different stator-to-rotor distances reveal the stators' magneto-mechanical contribution towards the mechanisms of damping.

The study shows that sinusoidal model for restoring torque as in (1) provides acceptable agreement with static and dynamic experiments for relatively large stator-to-rotor distances. However, for small stator-to-rotor distances, the sinusoidal model cannot even qualitatively capture the type of nonlinearity manifested in the rotor frequency sweep response. This response exhibits an intriguing stiffening-to-softening type of nonlinearity that requires developing a more advanced restoring torque model in later studies.

Since the decrease in stator-to-rotor distance provokes the transition from softening nonlinearity to stiffening-to-softening nonlinearity, interesting design considerations arise regarding the possibility to achieve some desired dynamic behaviors like a robust linear response, or an essential nonlinear response.

## 8. Acknowledgments

This work was supported by DARPA A MEchanically Based Antenna (AMEBA) program grant # DARPA HR0011-17-2-0057.

# Design, Dynamics, and Dissipation of a Torsional-Magnetic Spring Mechanism: Supplemental Material

Ali Kanj, Rhinithaa P. Thanalakshme, Chengzhang Li, John Kulikowski, Gaurav Bahl, Sameh Tawfick

This document contains the supplemental material to the work in "Design, Dynamics, and Dissipation of Torsional-Magnetic Spring Mechanism", which is referred to as the "main article". The document contains the sections of the following table of contents.

## Table of Contents

# S1. Instrumentation and Experiments

## A. Measurements Tower Setup

In order to proceed with the measurements of the magnetic-spring oscillator, the tower setup in Fig. S1 is implemented. This tower is equipped with the torque sensor (Fig. S1-c) and the fluxgate sensor (Fig. S1-d). Fig. S1-e shows the driving coil and its 3D printed PLA+ support.

For dynamic measurements of the device, the rotor needs to be isolated from the torque sensor that possesses considerable rotational inertia and damping. This isolation is possible by sliding the soft coupler in the configuration depicted in Fig. S1-f.

However, for quasi-static torque and angle measurements (such as the measurements in sections III and V in the main article), the soft coupler is attached to both the rotor's and the torque sensor's shafts. In these experiments, the rotor is rotated by a manual torque $T_{Manual}$ applied to the bottom side of the torque sensor's shaft as indicated in Fig. S1-b (see supplemental video 1). Slow manual rotation sustains a quasi-static equilibrium between the manual torque and the torques applied on the rotor PM. This quasi-static torque is uniform along the shaft of rotation (from bottom to the top of the tower in Fig. S1) and is measured by the torque sensor.

The following subsections explain how this tower setup is used along with the instruments in Fig. S2 in order to execute and measure the experiments of the main article.

## B. Static Experiments

A FUTEK TRS705 non-contact shaft to shaft rotary torque sensor with encoder is used for torque and angular displacement measurements (Fig. S1).

For static torque measurements in sections III (Fig. 2a) and V (Fig. 5a-b) of the main article, one end of the FUTEK sensor is connected to the rotor shaft by the mean of soft coupling (Fig. S1-f), the other end is manually slowly rotated. The slow rotation results in quasi-static equilibrium (i.e., without contribution of inertia and damping effects) between the manually applied torque from one side of the FUTEK sensor, and the torque applied at the rotor from the other of the sensor (see Fig. S1 and supplemental video 1). This torque applied on the rotor is induced by the magnetic spring ($T_{Mag}$) in section III, and by the coil ($T_{Coil}$) in section V. Thus, during the rotor rotation, the FUTEK sensor is measuring the applied torque on the rotor and the corresponding angular displacement.

During these measurements, these sensor recordings are continuously logged to the lab's computer by the mean of a NI USB-6210 DAQ (Fig. S2-a). The angular encoder of FUTEK sensor using the NI DAQ counter and LabVIEW "X4" decoding method allows to measure the angular displacement $\theta$ with a 0.25° resolution.

In the experiments of section V (Fig. 5c-d), the high current switching DC power supply BK Precision 1693 of Fig. S2-b delivers and measures the dc electric current ($i_{Coil}$) required to power the coil.

### C. Free Response Experiments

For free response experiments in the ringdowns of sections IV (Fig. 3b,c) and the free-full rotations of section V (Fig. 5c-d), the angular measurements are acquired by means of a Texas Instruments DRV425EVM integrated fluxgate magnetic sensor (Fig. S1). Located 19 cm from the rotor center and oriented along the y-axis of Fig. 1a, this sensor measures the change in the y-component of the magnetic field produced by the rotor's PM at the sensor's location.

The correspondence between the angular displacement and the voltage output of the magnetic sensor is determined by calibration done in subsection SI.E using the angular encoder of FUTEK TRS705 during quasi-static measurements (cf. Fig. S3). Tektronix DPO2014B oscilloscope (Fig. S2-d) captures the magnetic sensor voltage signal and consequently the evolution of $\theta$ as function of time during these measurements. In addition, the oscilloscope serves to measure the voltage drop across the coil terminals for the experiments in sections V, and VI.

For section V, the experimental angular velocity $\Omega = \frac{d\theta}{dt}$ are deduced by numerical differentiation using two-point stencil central differences method [1] in MATLAB of the measured angle time series.

### D. Frequency Sweep Experiments

In the frequency sweeps of section VI (Fig. 6 and 7), an HP 33120A function generator (Fig. S2-f) controls the frequency and the amplitude of the coil input voltage. The respective sinusoidal voltage created by the function generator is applied across the coil by the means of Gemini XGA-5000 professional power amplifier of Fig. S2-e (with 1:1 voltage gain) to overcome the maximum power limitation of the function generator signal. The electric current delivered to the coil is measured by a Hantek CC-65 AC/DC current clamp. The net magnetic field induced from this current flow and rotor oscillation is measured using the calibrated magnetic sensor.

The monitored electric current and magnetic field are transferred to two Stanford Research Systems SR860 lock-in amplifiers (Fig. S2-h) that admit the function generator signal as reference for frequency and phase shift. The lock-in amplifiers retrieve the phasor components (amplitude and phase shift) of the monitored electric current and magnetic field at the reference frequency.

A MATLAB script sweeps the frequency of the function generator, waits (~ 2 s) for steady state to be reached and then collects the phasor measurements of lock-in amplifiers. The MATLAB script runs this algorithm for the shown frequency range in Fig. 6b,c and 7 by increasing/decreasing the frequency in the forward/backward sweep.

To measure the angular displacement amplitudes $\theta_{amp}$ of Fig. 6 and 7 in the main article, the measured magnetic field is assumed to be the superposition of the coil and rotor magnetic fields. This superposition translates into the following phasor relation for every frequency component:

$$|B_{Net}| \angle \phi_{B_{Net}/i} = |B_{Rotor}| \angle \phi_{B_{Rotor}/i} + b_I \cdot i_{amp} \quad (s1)$$

where, respectively, for the considered frequency component:

- $i_{amp}$, $|B_{Rotor}|$ and $|B_{Net}|$ denote the amplitude of the electric current, the rotor magnetic field and the net magnetic field
- $\phi_{B_{Rotor}/i}$ and $\phi_{B_{Net}/i}$ denote the phase shift w.r.t. the electric current of the rotor magnetic field and the net magnetic field
- $b_I$ denotes the proportionality constant that scales the sensed magnetic field induced by the coil with the coil current.

Equation (s1) uses the linearity between the current passing through a coil and its induced magnetic field. This property is validated experimentally during the system-idnetification of the coil (cf. section SVI) by conducting the same frequency sweep but by fixing the rotor which identifies the proportionality constant $b_I$. Noting that $i_{amp}$, $|B_{Net}|$ and $\phi_{B_{Net}/i}$ of (s1) are retrieved by the simultaneous use of the two lock-in amplifiers, $|B_{Rotor}|$ can be computed from:

$$|B_{Rotor}| = \left[ \left( |B_{Net}| \cdot \cos \phi_{\frac{B_{Net}}{i}} - b_I \cdot i_{amp} \right)^2 + \left( |B_{Net}| \cdot \sin \phi_{B_{Net}/i} \right)^2 \right]^{\frac{1}{2}}. \quad (s2)$$

Finally, the oscillation amplitude $\theta_{amp}$ is directly deduced from $|B_{Rotor}|$ of (s2) by referring to the magnetic sensor calibration of subsection SI.E. By the mean of this procedure, the experimental plots of Fig. 6 and 7 are realized.

*E. Fluxgate Sensor Angular Calibration*

The angular displacements in dynamic experiments (i.e., sections IV, V, and VI in the main article) are measured via the fluxgate sensor (Fig. S1-b). For this purpose, the fluxgate sensor is calibrated for angular displacements using the FUTEK rotary torque sensor (Fig. S1-c). In the position of Fig. S1-a, the fluxgate measures the change in the magnetic field along the y-direction induced by the rotation of the rotor. This magnetic field change develops a voltage across the sensor. In order to relate the developed sensor voltage $V_B$ with the rotor angular displacement $\theta$, the rotor attached to the torque sensor is rotated while recording both the angular measurements of the FUTEK sensor and the fluxgate sensor. These simultaneous measurements produce the plot of Fig. S3 for the device with $d_{Stator} = 3.7$ cm. The experimental measurements are fit to a sinusoidal function of the form:

$$V_B = A \sin \theta + B \tag{s3}$$

where $A$ and $B$ are identified by the fit. It is experimentally found that the value of $A = 1.12$ V is constant for the same rotor and the same position of the sensor; both of which are fixed throughout the whole experiments. However, $B$ depends on the surrounding DC magnetic field like the presence of neighboring permanent magnets and the respective stator-to-rotor distance. Hence, an angular change $\Delta\theta$ is estimated by the change in voltage $\Delta V_B$ which eliminates the dependence on $B$ using the following relation:

$$\Delta\theta = \operatorname{asin}\frac{\Delta V_B}{A}. \tag{s4}$$

## S2. Derivation of Magnetic Spring Model

To model the effect of stator-to-rotor distance on the magnetic spring in the relation $T_{amp} = \alpha \times g(d_{Stator})$ of section III, the rotor is treated by its equivalent magnetic dipole moment ($\vec{m}_R$) and the stators magnetic field ($B_S$) is computed by the manufacturer's (K&J Magnetics) model. The manufacturer's model provides an expression for the magnetic field induced at the central line of a rectangular prismatic magnet [2] as function of the magnet's length $L$, width $W$, thickness $T$, residual flux density $B_r$ and the distance from its surface $d = d_{Stator} - 0.5T$. Using the expression in [2], equation (s5) is developed to estimate the stator magnetic field magnitude $B_S$ at the rotor center:

$$B_S = \frac{2B_r}{\pi} \times \left[ \arctan\left(\frac{L \times W}{2d\sqrt{4d^2 + L^2 + W^2}}\right) \right.$$
$$\left. - \arctan\left(\frac{L \times W}{2(d+T)\sqrt{4(d+T)^2 + L^2 + W^2}}\right) \right]. \tag{s5}$$

Substituting $B_S$ in (2) with (s5) results in the model in (s6) of $T_{amp}$ (which theoretically admits the same value as $K_{Mag}$):

$$T_{amp} = \underbrace{\frac{2m_R \cdot B_r}{\pi}}_{\alpha} \left[ \arctan\left(\frac{L \times W}{2d\sqrt{4d^2 + L^2 + W^2}}\right) \right.$$
$$\left. - \arctan\left(\frac{L \times W}{2(d+T)\sqrt{4(d+T)^2 + L^2 + W^2}}\right) \right]. \tag{s6}$$

The rotor equivalent magnetic moment magnitude $m_R$ is computed as [3]:

$$m_R = \frac{B_r}{\mu_0} \times Rotor\ PM\ Volume = 8.80 \times 10^{-3}\ \text{A} \cdot \text{m}^2. \tag{s7}$$

In (s7), $\mu_0$ is the magnetic permeability of a vacuum with a value of $4\pi \times 10^{-7}\ \text{N} \cdot \text{A}^{-2}$ [3]. The residual flux density $B_r$ that appears in (s5), (s6) and (s7) is a property of the magnetic material depending on its grade [3]. $B_r$ characterizes the remaining magnetic induction in the magnetic material after removal of external magnetization. For the Neodymium N45 magnets used in the studied device, a value of $B_r = 1.35$ T is used [4]. By the mean of (s7), a theoretical estimation of the pre-factor $\alpha$ of (s6) becomes:

$$\alpha = \frac{2}{\pi\mu_0} \times B_r^2 \times Rotor\ PM\ Volume = 7.57\ \text{N} \cdot \text{m}. \tag{s8}$$

Using the stators' actual dimensions ($L = 76.2$ mm and $W = T = 12.7$ mm), the fitting results in Fig. 2b show that this model captures the experimental scaling between magnetic spring stiffness and the stator-to-rotor distance with an $\alpha_{Fit} = 7.16\ \text{N} \cdot \text{m}$ which is with ~6% of its theoretical estimation in (s8).

## S3. Derivation of Ringdown Model with Combined Damping

The solution of the ringdown response with combined damping (i.e., equations (5) and (6) in the main article) is derived in this section. This derived solution expressing the angle $\theta$ as function of time $t$ which facilitates and speeds the fitting of the experimental data as explained by the end of this section.

Without loss of generality, the solution of (11) is derived by considering the following initial conditions:

$$\begin{cases} \theta(t = 0 \text{ s}) = \theta_0 \\ \dot{\theta}(t = 0 \text{ s}) = \Omega_0 \end{cases}. \tag{s9}$$

Following Den Hartog's approach in solving the forced response of a second order system with combined damping [5], the motion is fragmented into segments of sign constant rotor's velocity. These segments are the half-cycles during which the rotor oscillates towards ($\dot{\theta} < 0$) or away from ($\dot{\theta} > 0$) the maximum angle. In other words, each half-cycle corresponds to the motion between two successive extremums i.e. from a maximum to a minimum or vice versa. Within any of these half-cycles, equation (6) of the main article is a linear ordinary differential equation (O.D.E.) as stated in (s10):

$$\ddot{\theta} + 2\zeta\omega_n\dot{\theta} + \omega_n^2\theta = -S_k\omega_n^2\theta_f. \tag{s10}$$

$S_k$ in (s10) denotes the sign of the angular velocity ($S_k = \text{sgn }\dot{\theta} = \pm 1$) during the considered half-cycle indexed by $k \in \{0, 1, 2, 3, ...\}$. Thus, the entire rotor's response is governed by a sequence of the second order O.D.E. in (s10) with $S_k$ alternating between $+1$ and $-1$ from one half-cycle to the next one. Hence, the dry friction term on the right-hand-side of (s10) acts as a constant forcing term to the second order O.D.E. The general solution of (s10) is a linear superposition of a homogeneous and a particular solution reflecting the effect of viscous and dry damping respectively:

$$\theta_k(t) = \Theta_k . e^{-\zeta\omega_n(t-t_k)} \cos[\omega_d(t - t_k) + \phi_k] - S_k\theta_f. \tag{s11}$$

In (s11), $t_k$ denotes the instant at which the half-cycle of index $k$ starts i.e. the instant the angular velocity switches sign except for $t_0 = 0$ s marking the start of the motion. $\Theta_k$ and $\phi_k$ are constant coefficients determined by applying the initial conditions of the $k^{\text{th}}$ half-cycle. These initial conditions are obtained by imposing continuity of angular displacement and velocity between the end of the $(k-1)^{\text{th}}$ half-cycle and the beginning of the $k^{\text{th}}$ half-cycle. With the aim of computing $t_k$, $\Theta_k$, and $\phi_k$, the angular velocity within the $k^{\text{th}}$ half-cycle is derived in (s12) by means of (15):

$$\dot{\theta}_k(t) = -\omega_d\Theta_k e^{-\zeta\omega_n(t-t_k)} \left\{ \frac{\zeta}{\sqrt{1-\zeta^2}} \cos[\omega_d(t-t_k) + \phi_k] \right. \\ \left. + \sin[\omega_d(t-t_k) + \phi_k] \right\}. \tag{s12}$$

Starting with the initial half-cycle with index $k = 0$, the coefficients $\Theta_0$ and $\phi_0$ are calculated by solving the system of equations in (s13) using (s9), (s11), and (s12) with $t = t_0 = 0$ s and $S_0 = \text{sgn}\,\Omega_0$:

$$\begin{cases} \Theta_0 \cos \phi_0 = \theta_0 + S_0 \theta_f \\ -\omega_d \Theta_0 \left( \dfrac{\zeta}{\sqrt{1-\zeta^2}} \cos \phi_0 + \sin \phi_0 \right) = \Omega_0 \end{cases}. \tag{s13}$$

It should be mentioned that this half-cycle of $k = 0$ does not necessary starts from an extremum and therefore it might not match half of an oscillation.

For $k \in \{1, 2, 3, \dots\}$, the definition of the switching instants $t_k$ enforces the velocity initial condition $\dot{\theta}_k = 0$ rad·s$^{-1}$ at $t = t_k$, the beginning of the $k^{\text{th}}$ half-cycle. Based on this initial condition, equation (s12) yields to the solution of $\phi_k$ in (s14) up to a remainder integer multiple ($m_k \in \mathbb{Z}$) of $\pi$:

$$\phi_k = -\arctan\left(\frac{\zeta}{\sqrt{1-\zeta^2}}\right) + m_k \pi. \tag{s14}$$

However, $\arctan\left(\frac{\zeta}{\sqrt{1-\zeta^2}}\right)$ is confined to the interval $[0 \text{ rad}, \frac{\pi}{2} \text{ rad}]$ because $\zeta \geq 0$. Bounding $\phi_k$ to the interval $(-\pi \text{ rad}, +\pi \text{ rad}]$ forces $m_k$ to take the value either 0 or +1. The value of $m_k$ is dictated by the type of extremum achieved at $t = t_k$. For this reason, equation (s15) displays the value of $\theta_k$ by evaluating (s11) at $t = t_k$:

$$\theta_k(t = t_k) = \Theta_k \cos \phi_k - S_k \theta_f. \tag{s15}$$

Considering the case of weak viscous damping ($\zeta \ll 1$), $\phi_k$ in (s14) is sufficiently close to 0 rad ($m_k = 0$) or $\pi$ rad ($m_k = 1$) so that $\cos \phi_k$ admits a value close to $\cos 0 = 1$ or $\cos \pi = -1$. In addition, the sign of $\theta_k$ at $t = t_k$ in (s15) is equal to that of $\Theta_k \cos \phi_k$ and is not affected by the $\theta_f$ term whose magnitude is supposed to be sufficiently small. Hence, with $\Theta_k \geq 0$, if the $k^{\text{th}}$ half-cycle starts from a maximum (/minimum) angle i.e. with positive (/negative) sign, then $m_k = 0$ (/$m_k = 1$) leading to $\cos \phi_k = +1$ (/$\cos \phi_k = -1$). Consequently, from (s14), $\phi_k$ is related to the damping ratio $\zeta$ and the velocity sign $S_k$ by (s16):

$$\phi_k = -\arctan\left(\frac{\zeta}{\sqrt{1-\zeta^2}}\right) + (1 + S_k)\frac{\pi}{2}. \tag{s16}$$

Equation (s16) takes advantage of the fact that $S_k = -1$ ($S_k = +1$) if the half-cycle starts from a maximum (/minimum). Note that $\phi_k$ can be recursively computed in (s17) for $k \in \{2, 3, \dots\}$ by using (s16) and the fact that $S_k = -S_{k-1}$:

$$\phi_k = \phi_{k-1} - S_{k-1}\pi. \tag{s17}$$

Furthermore, $t_k$ being the instant at which the motion changes direction, then it is identified in (s18) using (s12) by setting $\dot{\theta}_{k-1}(t = t_k) = 0$:

$$t_k - t_{k-1} = -\frac{1}{\omega_d}\left[\arctan\left(\frac{\zeta}{\sqrt{1-\zeta^2}}\right) + z_{k-1}\pi + \phi_{k-1}\right]. \tag{s18}$$

From (s18) where $z_{k-1} \in \mathbb{Z}$, (s17) allows to deduce (s19):

$$t_k - t_{k-1} = -\left(\frac{1+S_{k-1}}{2} + z_{k-1}\right)\frac{\pi}{\omega_d}. \tag{s19}$$

Nevertheless, for $t \in (t_{k-1}, t_k]$, $\dot{\theta}_{k-1}$ achieves the zero value exactly once. Thus, for $t \in (t_{k-1}, t_k]$, a unique value of $z_{k-1} \in \mathbb{Z}$ satisfies (s19) as displayed by (s20):

$$\begin{cases} S_{k-1} = -1 \Rightarrow z_{k-1} = -1 \\ S_{k-1} = +1 \Rightarrow z_{k-1} = -2 \end{cases} \Rightarrow t_k - t_{k-1} = \frac{\pi}{\omega_d}. \tag{s20}$$

In summary, the periods of half-cycles for $k \geq 1$ are equal to half of the period ($\frac{1}{2}T_d$) of damped oscillations with exclusively viscous dissipation (Fig. 3a) i.e. dry friction does not alter the frequency of viscous damped oscillations. Hence, we can infer from equation (s20) that the instants $t_k$ form an arithmetic sequence that leads to (s21) for $k \in \{1, 2, 3, \dots\}$:

$$t_k = t_1 + (k-1)\frac{\pi}{\omega_d}. \tag{s21}$$

The initial velocity switching instant $t_1$ is obtained in (s22) by also applying (s17) but with the value of $\phi_0$ identified from (s13) and with $t_0 = 0$ s:

$$t_1 = -\frac{1}{\omega_d}\left[\arctan\left(\frac{\zeta}{\sqrt{1-\zeta^2}}\right) + z_0\pi + \phi_0\right]. \tag{s22}$$

Similar to $z_{k-1}$ in (s19), $z_0$ in (s22) is the unique integer in $\mathbb{Z}$ that results in the smallest strictly positive $t_1$. Hence, this property of $z_0$ in (s22) leads to (s23) where the floor function operates on a real number to return the preceding integer:

$$z_0 = \text{floor}\left[-\frac{1}{\pi}\left(\arctan\frac{\zeta}{\sqrt{1-\zeta^2}} + \phi_0\right)\right]. \tag{s23}$$

Knowing that $\left(\arctan\frac{\zeta}{\sqrt{1-\zeta^2}} + \phi_0\right) \in \left(-\pi, \frac{3\pi}{2}\right]$ rad then $z_0 \in \{-2, -1, 0\}$. With $t_1$ known from (s22), the arithmetic sequence (s20) allows to identify in (s24) the index $k$ of the half-cycle as function of time $t$:

$$\begin{cases} \text{If } t < t_1: k = 0 \\ \text{If } t \geq t_1: k = \text{floor}\left[\dfrac{\omega_d(t - t_1)}{\pi}\right]. \end{cases} \tag{s24}$$

Finally, the strictly positive coefficients $\Theta_k$ for $k \in \{1, 2, 3, \dots\}$ are determined by requiring $\theta_{k-1}(t = t_k) = \theta_k(t = t_k)$. This motion continuity condition allows to write the recursive relation in (s25) for $k \in \{2, 3, \dots\}$ when applied to (s11) with the use of (s16), (s17), (s20), and that $S_{k-1} = -S_k$:

$$\Theta_k = \Theta_{k-1} e^{-\pi\zeta/\sqrt{1-\zeta^2}} - \dfrac{2\theta_f}{\sqrt{1-\zeta^2}}. \tag{s25}$$

Equation (s25) make use of the property deduced from (s16) that $\cos\phi_k = -S_k\sqrt{1-\zeta^2}$. Equation (s25) details the form of amplitudes' decrement exerted by the combined dissipations. As expected, the decrement is a mix of logarithmic (proportional) and linear attenuations that recuperates the same decrements discussed earlier with individual type of damping (Fig. 3a,b). Furthermore, for $k \in \{2, 3, \dots\}$, an explicit formula of $\Theta_k$ in (s26) is concluded from (s25) by mathematical induction and then applying the sum identity of geometric series:

$$\begin{aligned}
\Theta_k &= \Theta_1 e^{-(k-1)\pi\zeta/\sqrt{1-\zeta^2}} - \dfrac{2\theta_f}{\sqrt{1-\zeta^2}} \sum_{n=0}^{k-2} e^{-\dfrac{n\pi\zeta}{\sqrt{1-\zeta^2}}} \\
&= \Theta_1 e^{-\dfrac{(k-1)\pi\zeta}{\sqrt{1-\zeta^2}}} - \dfrac{2\theta_f}{\sqrt{1-\zeta^2}} \left[\dfrac{1 - e^{-\dfrac{(k-1)\pi\zeta}{\sqrt{1-\zeta^2}}}}{1 - e^{-\dfrac{\pi\zeta}{\sqrt{1-\zeta^2}}}}\right].
\end{aligned} \tag{s26}$$

However, for $k = 1$, continuity of motion that resulted in (s25) for $k \in \{2, 3, \dots\}$ calculates $\Theta_1$ in (s27) because (s20) does not apply:

$$\Theta_1 = \Theta_0 \dfrac{e^{-\zeta\omega_n t_1} \cos(\omega_d t_1 + \phi_0)}{S_0\sqrt{1-\zeta^2}} - \dfrac{2\theta_f}{\sqrt{1-\zeta^2}} \tag{s27}$$

For instance, $S_0 = \text{sgn}\,\Omega_0$ is related to $\phi_0$ and $\zeta$ by (s28) arising from the second equation in (s13):

$$S_0 = -\text{sgn}\left(\dfrac{\zeta}{\sqrt{1-\zeta^2}}\cos\phi_0 + \sin\phi_0\right). \tag{s28}$$

Hence, equation (s28) permits to compute the value of $S_k$ for $k \in \{1, 2, 3, \dots\}$ in (s29) that is derived due to the alternating nature of $S_k$:

$$S_k = (-1)^k \times S_0. \tag{s29}$$

To summarize, the derived equations in this subsection explicitly predict the instantaneous angular displacement $\theta(t)$ during ringdown with combined dissipations for a given set of parameters: $\{\omega_n, \zeta, \theta_f, \Theta_0, \phi_0\}$. This set of parameters – to be treated as fitting parameters for the experimental time series $\theta(t)$ – allows to compute $z_0$, $t_1$, $S_0$, and $\Theta_1$ by applying (s23), (s22), (s28), and (s27) respectively (in that order). After computing these coefficients that characterize the end of the starting half-cycle, in order to calculate the angular displacement $\theta$ during combined damping ringdown, at every instant of time $t$:

i. The index $k$ of the current half-cycle is computed in (s24)
ii. This value of $k$ calculates $t_k$, $\Theta_k$, and $S_k$ using (s21), (s26), and (s29) respectively
iii. $S_k$ is applied in (s16) to compute $\phi_k$
iv. The angular displacement $\theta$ at the considered instant $t$ is calculated by (s11) where $\omega_d = \omega_n\sqrt{1-\zeta^2}$.

This method of analytically solving for $\theta$ at every instant $t$ is referred to as the combined damping model in section IV. To model the combined damping in equation (8) of the main article, equations (s16), (s21), (s22), (s23), (s26), (s27), and (s29) are applied in (s11) knowing by induction that $\cos[\omega_d(t - t_k) + \phi_k] = \cos[\omega_d t + \phi]$ using equations (s17), (s19), (s22), and (s23).

*Numerical Identification of Ringdown Models*

Applying the analytical explicit $\theta(t)$ models of (7) and (8) is computationally much less expensive than identifying damping with the system's equation of motion (6) for computing angular displacement. Essentially, simulating the evolution of $\theta$ between two sampling instants (e.g. from $\hat{t}_1$ to $\hat{t}_2$) with an acceptable accuracy level requires a sufficiently refined simulation time step (i.e. smaller than $\hat{t}_2 - \hat{t}_1$). This criterion means that, unless the duration between the two sampling instants is sufficiently small, additional computations should be run between the two considered data points in order to achieve a certain accuracy. On the other hand, the implemented analytical model solves for the exact values directly at the considered sampling instants. The importance of this computational efficiency is crucial because the model is run multiple times for every conducted fitting.

As for the experimental data, the collected time series are trimmed to the time segment where the oscillations' amplitude decays from 15° to 2.5° (cf. Fig. 3c and 3d). This trimming limits the nonlinear behavior of the magnetic spring. The trimmed measurements are filtered by a Butterworth low-pass-filter of order five with a cutoff frequency equal to hundred times the oscillations frequency obtained by an FFT. The filtered and trimmed data is well fitted to the derived analytical solutions.

Therefore, fitting experimental data to the combined damping model results in a set of five parameters: $\{\omega_n, \zeta, \theta_f, \Theta, \phi\}$ whereas the solely-viscous damping model results in a set of four parameters: $\{\omega_n, \zeta, \Theta, \phi\}$. The fitting uses a MATLAB function called "MultiStart" included in the global optimization toolbox. MultiStart searches for the set of parameters, respective to each model, that leads to an absolute (global) minimum of the sum of square-errors computed by "lsqcurvefit" function.

## S4. Energy Evolution Metric for Comparison of Damping Models in Ringdowns

In Fig. 4d of section V in the main article, the actual (experimental) energy dissipated ($E_{Dis}$) in at each instant ($t$) is computed by subtracting the instantaneous mechanical energy (kinetic and elastic) from the initial mechanical energy of the oscillator as dictated by:

$$E_{Dis}(t) = E_{Mech}(t=0) - E_{Mech}(t) \tag{s31}$$

The mechanical energy of the oscillator comprises the elastic energy ($E_{Elas}$) in the magnetic spring and the kinetic energy ($E_{Elas}$) of the rotor as given by (s32) at each time instant:

$$E_{Mech}(t) = \underbrace{\frac{1}{2} K_{Mag} \cdot \theta(t)^2}_{E_{Elas}(t)} + \underbrace{\frac{1}{2} J \cdot \dot{\theta}(t)^2}_{E_{Kin}(t)}. \tag{s32}$$

The values of $K_{Mag}$ and $J$ used in (s32) are the ones experimentally identified in Fig. 4a (precisely $K_{Mag}$ from static measurements). The rotor instantaneous velocity $\dot{\theta}(t)$ is calculated by numerical differentiation using height-point stencil central difference method [1] in MATLAB.

Further, each damping model estimates the total energy dissipation by:

$$E_{Dis}(t) = \int_{\text{Initial State}}^{\text{State at } t} dW_{Dis}. \tag{s33}$$

In (s33), $dW_{Dis}$ is the dissipative work done for an infinitesimal angle change of $d\theta$ by the viscous damping torque $T_{Vis}$ in (7) and dry friction torque $T_{Dry}$ in (8) as stated by (s34):

$$dW_{Dis} = c\dot{\theta}d\theta + T_f|d\theta|. \tag{s34}$$

The integration of $dW_{Dis}$ in (s33) is done numerically using trapezoidal method in MATLAB

and leads to results similar to the ones in Fig. 4d for $d_{Stator} = 3.1$ cm. However, to evaluate the performance over all measured $d_{Stator}$, the inset of Fig. 4d presents the relative root-mean-square (RMS) error defined as:

$$Relative\ RMS\ Error = \frac{[\sum_{t_i} \Delta E_{Dis}(t_i)^2]^{\frac{1}{2}}}{\max(E_{Dis}|_{Exp})}. \tag{s35}$$

In (s35), the square of difference $\Delta E_{Dis}(t_i)$ between the experimental and modeled energy dissipation is summed over all the sampled instants $t_i$. This sum is normalized by the maximum experimental energy dissipated $\max(E_{Dis}|_{Exp})$ which is the total energy dissipated at the end of oscillations equivalently the initial energy of the oscillator.

## S5. Derivation of Electromechanical Coupling Models

### A. Derivation of Torque Coupling Model

To develop reduced order models of the electromechanical coupling, the rotor is, once more, approximated by its equivalent magnetic dipole moment $\vec{m}_R$. The driving torque $T_{Coil}$ of Fig. 5a applied on the rotor magnetic dipole moment $\vec{m}_R$, under the effect of the coil magnetic field at the rotor's center $\vec{B}_{Coil} = B_{Coil} \cdot \hat{\jmath}$, can be computed by [6]:

$$\vec{T}_{Coil} = \vec{m}_R \times \vec{B}_{Coil} = (m_R \cdot B_{Coil} \cdot \cos\theta)\,\hat{k} \Rightarrow T_{Coil} = m_R \cdot B_{Coil} \cdot \cos\theta. \tag{s36}$$

According to Biot-Savart law, the coil magnetic field $B_{Coil}$ is proportional to the current $i_{Coil}$ [6] which allows to express $T_{Coil} = k_T \times i_{Coil} \times \cos\theta$ as in equation (7) of the main article where the torque coupling coefficient $k_T$ accounts for the proportionality constant relating $B_{Coil}$ to $i_{Coil}$ in addition to the magnitude of the rotor dipole moment $m_R$.

### B. Derivation of Back-Electromotive Force Coupling Model

The back-electromotive force $u_{EMF}$ induced across the coil results from the change in its magnetic flux $\phi_{Coil}$ due to the nearby rotor's PM movement as illustrated by Fig. 5c. According to Faraday's law, $u_{EMF}$ opposes (in right-hand-rule sense) the rate of change in coil magnetic flux as in (s37) where the dot operator denotes time differentiation [6]:

$$u_{EMF} = -\dot{\phi}_{Coil}. \tag{s37}$$

For the sake of implementing a simple electromechanical coupling model, the coil is approximated by its equivalent current loop of aggregate surface area $A_{Coil}$ that is located at the coil center at a distance $d_{Coil}$ from the rotor. Moreover, the magnetic field of the rotor $\vec{B}_{Rotor}$ is assumed to be uniform within the equivalent current loop. Under these assumptions, the coil magnetic flux $\phi_{Coil}|_{Rotor}$ resulting from the rotor can be expressed as in (s38):

$$\phi_{Coil}|_{Rotor} = A_{Coil}(\vec{B}_{Rotor} \cdot \hat{j}) . \tag{s38}$$

In addition, the rotor's uniform magnetic field through the current loop $\vec{B}_{Rotor}$ is assumed to be equal to the magnetic field at the loop center. Along with previous assumptions, expressing the dipole of the rotor as $\vec{m}_R = m_R(\hat{i}\cos\theta + \hat{j}\sin\theta)$ allows to compute $\vec{B}_{Rotor}$ as given by (s39) [6]:

$$\vec{B}_{Rotor} = \frac{\mu_0}{4\pi} \times \frac{m_R}{d_{Coil}^3} \times (-\hat{i}\cos\theta + 2\hat{j}\sin\theta) . \tag{s39}$$

With the use of (s38) and (s39), and denoting the rotor angular speed by $\Omega = \dot{\theta}$, the back electromotive force of (s37) is modeled as shown in (s40):

$$u_{EMF} = -\underbrace{\frac{\mu_0}{2\pi} \times \frac{m_R}{d_{Coil}^3} \times A_{Coil}}_{k_{EMF}} \times \Omega \times \cos\theta . \tag{s40}$$

The back-EMF coupling coefficient $k_{EMF}$ in (s40) groups the effect of the rotor magnetic field at the coil center and the total area of the coil windings. This $k_{EMF}$ model provides the dependence $k_{EMF} = \beta d_{Coil}^{-3}$ with:

$$\beta = \frac{\mu_0}{2\pi} \times m_R \times A_{Coil} . \tag{s41}$$

A simple way to approximate $\beta$ is to express (s41) in terms of the coil self-inductance $L$. With the same model assumption, the coil compressed into an equivalent current loop of aggregate surface area $A_{Coil}$ admits an inductance $L$ such that:

$$L = \frac{\mu_0}{2\pi R_{Loop}} A_{Coil} \Rightarrow A_{Coil} = \frac{2\pi R_{Coil}}{\mu_0} L . \tag{s42}$$

Equation (s42) provides an estimate of the assumed current loop effective surface area $A_{Coil}$ as function of $\mu_0$, $L$, and the current loop radius $R_{Loop}$. In order to compute $A_{Coil}$ in (s42), $R_{Loop}$ is taken to be the average radius (~18 mm) of the actual coil of Fig. 1c and the identified $L$ of Table I (as explained in the section SVI) is used. By the mean of these values, (s7), (s41), and (s42), $\beta$ is estimated to be ~$2.90 \times 10^{-7}$ V·s·m$^3$·rad$^{-1}$ which lies within 15% of the value found from fitting experimental data ($\beta_{Fit} \approx 3.32 \times 10^{-7}$ V·s·m$^3$·rad$^{-1}$).

## C. Conservative Electromechanical Conversion and Coefficients Equality

The equality between the two experimental electromechanical coupling coefficients ($k_T$ and $k_{EMF}$) in Fig. 5e indicates a conservative (lossless) electrical and mechanical conversion through the magnetic coupling.

Actually, based on the model in (47), the coil delivers an instantaneous mechanical power to the rotor as stated in (s43):

$$P_{Mech} = T_{Coil} \times \Omega = k_T \times (\Omega . i_{Coil} . \cos\theta) . \tag{s43}$$

Likewise, based on (s40), the instantaneous electrical power delivered by the rotor to the coil is calculated in (s44):

$$P_{Elec} = u_{EMF} \times i_{Coil} = k_{EMF} \times (\Omega . i_{Coil} . \cos\theta) . \tag{s44}$$

Hence, in order for the magnetic coupling system to convert the received electrical power totally into mechanical one and vice versa (i.e., $P_{Mech} = P_{Elec}$), Equations (s43) and (s44) require the equality between $k_T$ and $k_{EMF}$ similar to the experimental results.

## S6. System-Identification of The Coil

For the frequency sweep study conducted in section VI of the main article, the following coil parameters need to be identified: $b_I$ in (s1), the values of resistance $R$ and the self-inductance $L$ in TABLE I.

### A. Identification of The Coil Magnetic Coefficient $b_I$

To identify $b_I$, the rotor is mechanically fixed, and the coil is driven by a sinusoidal electric signal. During this electric drive, the current flowing through the coil is measured via the Hantek CC-65 AC/DC current clamp and the induced magnetic field via the fluxgate sensor (Fig. S1-d), and their amplitudes are recorded simultaneously via the two lock-in amplifiers.

The resulting steady state amplitudes of the measured current ($i_{amp}$) and magnetic field ($B_{Coil}$) are reported in Fig. S4 for two excitation frequencies (60 Hz and 100 Hz) for three different stator-to-rotor distances. The experimental datapoints of Fig. S4 depict that a proportionality relation governs $B_{Coil}$ to $i_{amp}$ as stated in (s1). A linear fit of the experimental datapoints with $d_{Stator} = 3.1$ cm results in $b_I = 4.07 \times 10^{-6}$ T/A whose value is applied for the study in section VI. The dashed line plotted in Fig. S4 demonstrates that this value of $b_I = 4.07 \times 10^{-6}$ T/A is appropriate for all the conducted measurements.

## B. Identification of the Coil Electric Coefficients R and L

As for identifying the coil resistance $R$ and self-inductance $L$, also while keeping the rotor mechanically fixed, the coil is subjected to frequency sweeps similar to the sweeps described in section SI.D with free rotor. In these sweeps, the coil's voltage and current are measured by the mean of the same setup described in section SI.D that uses the two lock-in amplifiers.

Hence, the amplitude of the electrical impedance $|Z|$ of the coil system (with fixed rotor) is computed for the different excitation frequencies $f$, and the experimental datapoints are presented in Fig. S5. The voltage across coil system is governed by:

$$U_{Coil} = L\frac{di_{Coil}}{dt} + Ri_{Coil} \,. \tag{s45}$$

By imposing a sinusoidal voltage signal of frequency $f$ across the coil, the steady state impendence of the system in (s45) becomes:

$$Z = \frac{U_{Coil}}{i_{Coil}} = j(2\pi f)L + R \,. \tag{s46}$$

Taking the amplitude of $Z$ in (s46) and raising its square provides the linear dependence between $|Z|^2$ and $f^2$ that is illustrated by the experimental datapoints of Fig. S5. A linear fit is applied for these experimental datapoints for every root-mean-square (RMS) value of the imposed coil voltage as shown in Fig. S5. Every fit provides a set of $R$ and $L$. Since, in the study of section VI, the voltage across the coil doesn't exceed 2 V in amplitude (Fig. 6b,c), the averages of $R$ and $L$ identified from the frequency sweeps with 1 V and 2 V RMS voltage (Fig. S5) are assigned for the coil as listed in TABLE I.

# S7. Supporting Figures

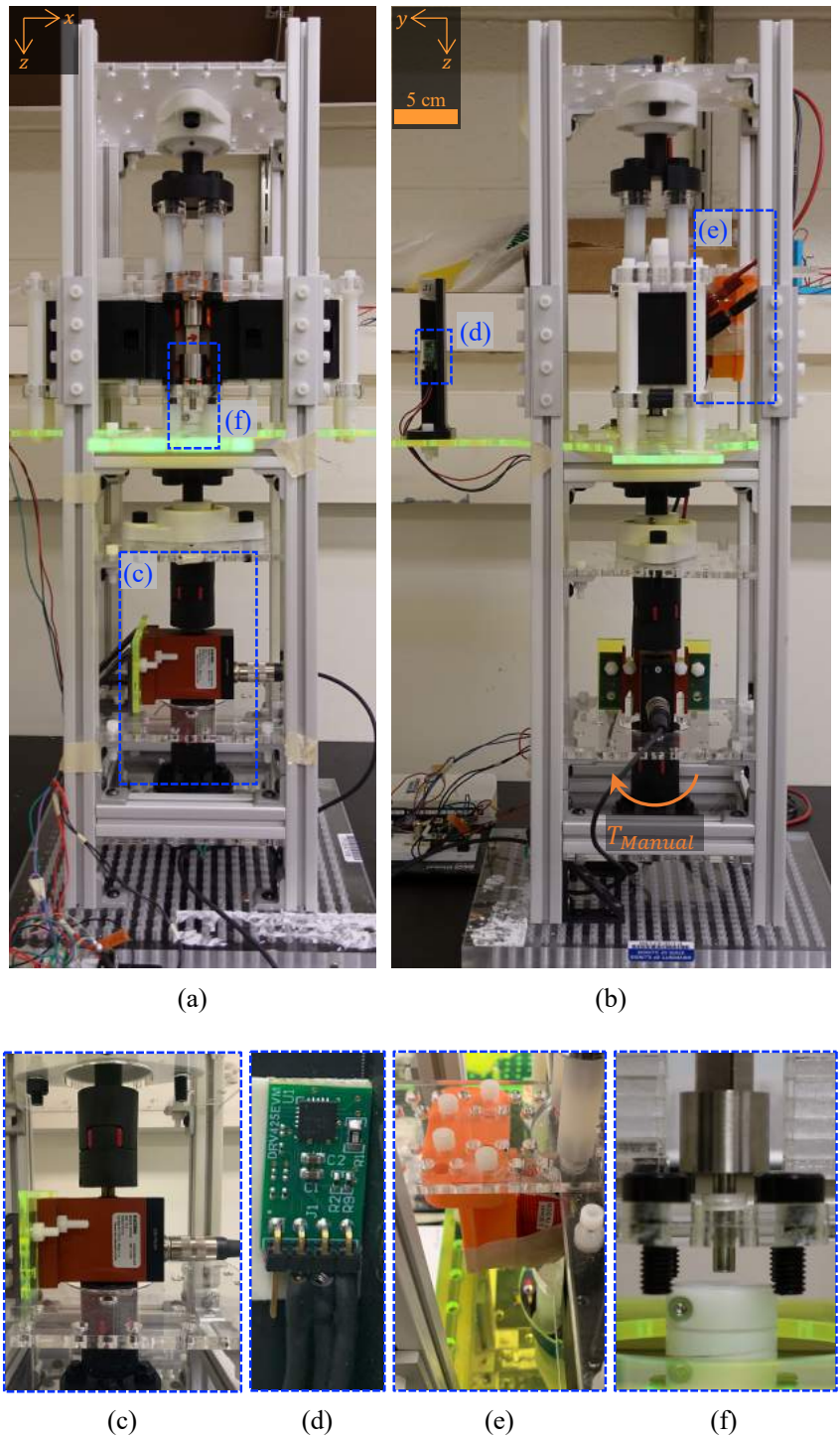

(a)　　　　　　　　　　　(b)

(c)　　　(d)　　　(e)　　　(f)

Fig. S1. Photographs of implemented tower setup used to measure magnetic torque and angular displacements in (a) front view and (b) side view with zoomed views over: (c) the FUTEK TRS705 rotary torque sensor, (d) the Texas Instruments DRV425EVM integrated fluxgate magnetic sensor, (e) the 18 AWG air-core inductor attached to its 3D-printed orange PLA+ support, and (f) the soft coupler when the rotor shaft is detached from the torque sensor.

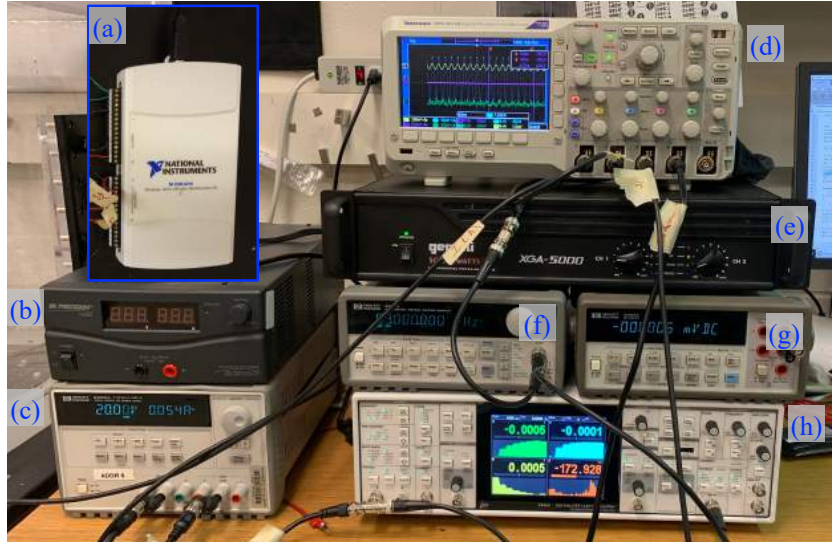

Fig. S2. Photo of instruments used in the experiments: (a) NI USB-6210 DAQ, (b) BK Precision 1693 high current switching DC power supply, (c) HP E3631A DC power supply, (d) Tektronix DPO2014B oscilloscope, (e) Gemini XGA-5000 professional power amplifier, (f) HP 33120A function generator, (g) HP 34401A multimeter, and (h) one of the two used Stanford Research Systems SR860 lock-in amplifiers.

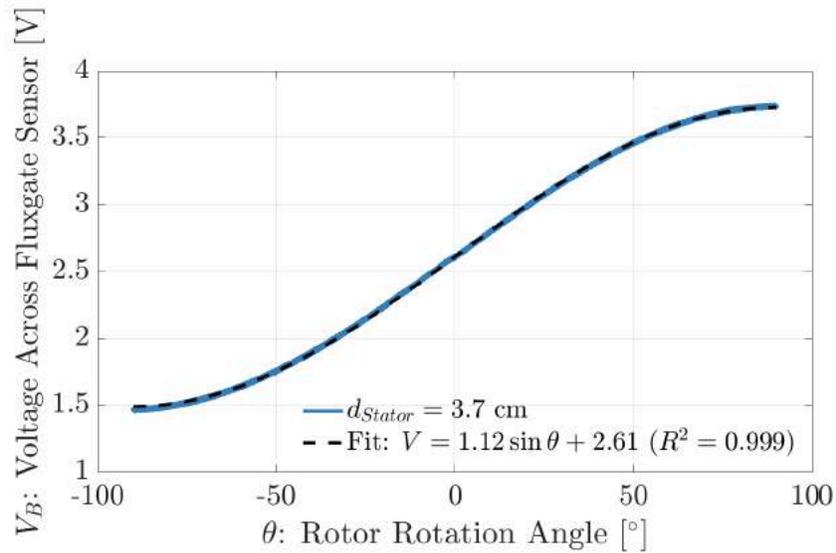

Fig. S3. Angular calibration of fluxgate sensor by measuring its voltage variation due to manual rotation of the rotor.

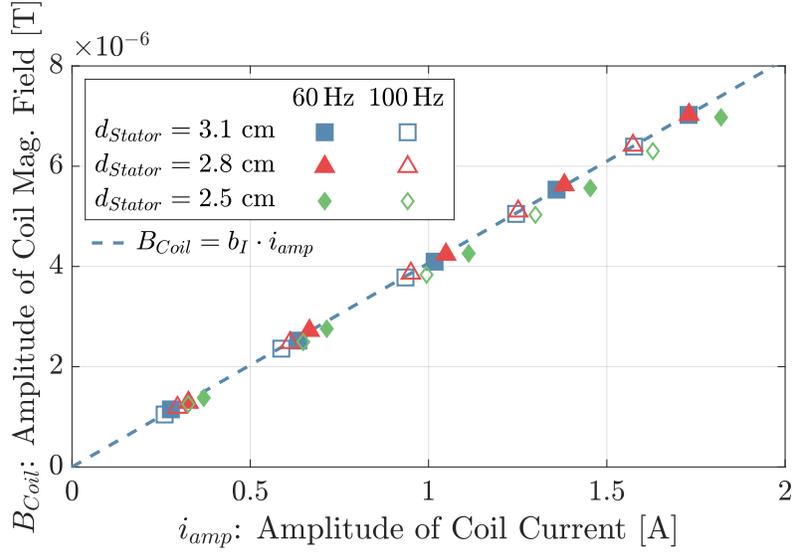

Fig. S4. Amplitude of the steady state magnetic field induced by the coil (only) due to sinusoidal current flowing through the coil for different stator-to-rotor distances ($d_{Stator}$) and excitation frequencies (60 Hz and 100 Hz). Dashed line corresponds to the linear correspondence used in (s1) with the identified $b_I = 4.07 \times 10^{-6}$ T/A from measurements with $d_{Stator} = 3.1$ cm.

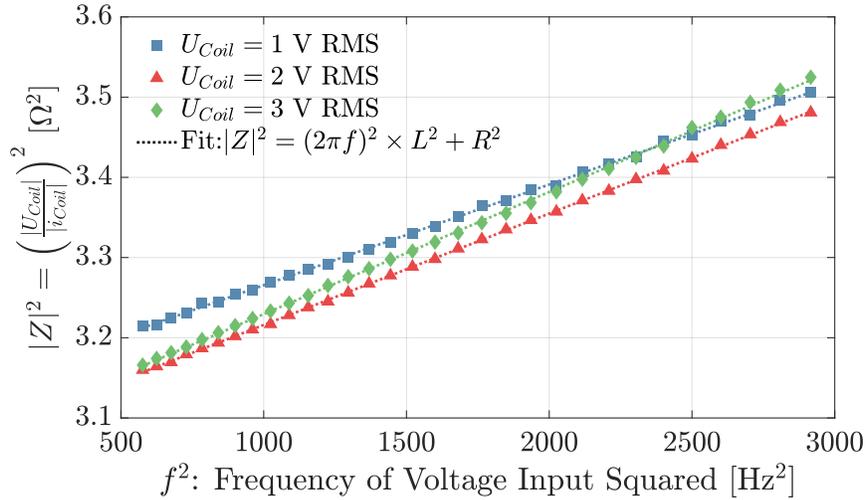

Fig. S5. Frequency sweep response of the coil impedance used to identify the coil resistance ($R$) and inductance ($L$) for different applied voltages in the system with stator-to-rotor distance $d_{Stator} = 3.4$ cm while fixing the rotor.

## S8. Supporting Videos

Supporting Video 1 shows a static torque $T_{Mag}$ measurement where the rotor is quasi-statically (i.e., slowly) rotated by applying a manual torque $T_{Manual}$ from beneath in the tower measurement setup (cf. Fig S1). In the Video 1, the manual torque $T_{Manual}$, the induced angle of rotation $\theta$, and the restoring magnetic torque $T_{Mag}$ of the stators are annotated.

Supporting Video 2 shows a forced response of the rotor in the device of $d_{Stator} = 3.4$ cm and $d_{Coil} = 3.4$ cm with large response amplitude for a frequency close to resonance.